\documentclass[journal=acsodf,manuscript=article]{achemso}
\usepackage{subcaption}
\usepackage{multicol}
\usepackage{float}
\usepackage{siunitx}
\usepackage{nomencl}
\usepackage{amsmath,amssymb,amstext,bm,array,multirow,booktabs,physics} % Lots of math symbols and environments
\usepackage{mathtools}
\usepackage{graphicx} % For including graphics N.B. pdftex graphics driver
\graphicspath{{figures/}}
\usepackage[capitalize]{cleveref}
\crefname{equation}{Eqn.}{Eqns.}    % get cleveref to use eqn instead of eq as the abbreviation for equation
\Crefname{equation}{Equation}{Equations}
\DeclarePairedDelimiterX{\inp}[2]{\langle}{\rangle}{#1, #2}

\makenomenclature

\title{Diffuse-interface blended method for imposing physical boundaries in two-fluid flows}

\author{Tanyakarn Treeratanaphitak}
\affiliation[SIIT]{School of Integrated Science and Innovation, Sirindhorn International Institute of Technology, Thammasat University, Pathum Thani 12121, Thailand}
\email{tanyakarn@siit.tu.ac.th}
\author{Nasser Mohieddin Abukhdeir}
\affiliation[UW]{Department of Chemical Engineering, University of Waterloo, 200 University Avenue West Waterloo, N2L 3G1, ON, Canada}
\alsoaffiliation[UWP]{Department of Physics \& Astronomy, University of Waterloo, 200 University Avenue West Waterloo, N2L 3G1, ON, Canada}

\begin{document}
%\nobibliography*       % uncomment this when using bibentry
%\listoftodos    % TODO list, comment out when finished

\begin{abstract}
Multiphase flows are commonly found in chemical engineering processes such as distillation columns, bubble columns, fluidized beds and heat exchangers.
The physical boundaries of domains in numerical simulations of multiphase flows are generally defined by a conformal unstructured mesh which, depending on the complexity of the physical system, results in time-consuming mesh generation which frequently requires user-intervention.
Furthermore, the resulting conformal unstructured mesh could potentially contain a large number of skewed elements, which is undesirable for numerical stability and accuracy.
The diffuse-interface approach allows for the use of a simple structured meshes to be used while still capturing the desired physical (\textit{e.g.} solid-fluid) boundaries.
In this work, a novel diffuse-interface method for the imposition of physical boundaries is developed for the incompressible two-fluid multiphase flow model.
This model is appropriate for dispersed multiphase flows which are pervasive in chemical engineering processes, in that this flow regime results in high levels of mass and energy transfer between phases.
A diffuse-interface is used to define the physical boundaries and boundary conditions are imposed by blending the conservation equations from the two-fluid model with that of the non-deformable solid.
The results from the diffuse-interface method are compared with results from a conformal unstructured mesh for different interface functions and widths.
For small interface widths, the accuracy of the flow profile is unaffected by the choice of interface function and the phase fraction distribution and flow behavior are within 3\% compared to those from a conformal mesh.
As the interface width increases, the diffuse-interface solution deviates from the conformal mesh solution in both the localized gas fraction and the overall gas hold-up, resulting in a difference up to 30\%.
In the case of flow past a cylinder, where the solid interacts with the flow, the presence of the diffuse-interface extends the thickness of the solid boundary and results in a deviation from the conformal mesh solution as time increases.
\end{abstract}

\printnomenclature

\section{Introduction}

Industrial chemical engineering processes such as bubble columns \citep{Jakobsen2005,Joshi2001,Ekambara2005,Krishna2001}, reactors \citep{Becker1994,Sokolichin1994}, pipe flow \cite{Ejaz2022,Ejaz2022a,Rasheed2022} and separators \citep{Lane2016a} involve multiphase flows which pose significant challenges for simulation-based design and optimization.
However, to improve existing designs and develop next-generation multiphase flow-based processes, an understanding of the complex hydrodynamics of the system is essential.
Increasingly the use of computational fluid dynamics (CFD) simulations are being used to study multiphase flow systems, enabling design and optimization activities that are infeasible using solely experimentation and physical prototyping.
CFD simulations of multiphase flow systems enable researchers to explore different combinations of operating conditions and prototype designs without the cost and safety issues incurred by experimental methods.

A vital aspect of the use of CFD simulations for the design and optimization of process equipment is the specification of internal physical features, which may have highly complex shape and topology.
These features need to be specified as physical boundaries in the simulation, which can be achieved by either using a conformal unstructured mesh or an embedded domain method.
With a conformal unstructured mesh, the geometry is defined such that once generated, the mesh surfaces correspond to the physical boundaries.
This process can be tedious, time-consuming and have detrimental numerical effects on the computational complexity and numerical stability of simulations, especially for complex geometries present in chemical engineering processes.
Additionally, if the internal features are changed, which is likely the case during design and optimization activities, the mesh will also have to change, thus requiring the mesh to be regenerated.
In the case of moving mesh problems, methods like the arbitrary Lagrangian-Eulerian (ALE) method \citep{Donea2004} is used, but ALE requires the mesh to be deformed as the boundary moves.

Instead of using a conformal mesh, the physical boundaries may be ``embedded'' in the problem, which has been a topic of research in the area of single-phase fluid mechanics for several decades, especially for fluid-structure interaction problems.
Examples of past relevant work includes the use of the fictitious domain \citep{Glowinski1999}, immersed boundary \citep{Mittal2005,Sotiropoulos2014,Griffith2020} and diffuse domain/interface \citep{Ramiere2007,Li2009b,Aland2010,Schlottbom2016,Nguyen2018,Monte2022} methods.
Physical boundaries are defined in the embedded domain method through the use of a level-set function, a phase-field, or similar continuous indicator field.
Since physical boundaries are not explicitly defined by the domain mesh, the mesh is not required to conform to them and a simple nonconforming structured mesh may be used.
This has many benefits, including the reduction of the need for remeshing when the geometry changes, along with improved numerical stability.
The ease with which the internal features can be evolved during simulation is highly beneficial for design and optimizing activities where the indicator field can directly be modified by a higher-level optimization scheme.

Focusing on the immersed boundary (IB) method, it has been extensively used to impose solid boundaries in single-phase flow.
Single-phase immersed boundary studies are reviewed in \citet{Mittal2005,Sotiropoulos2014} and \citet{Griffith2020}.
The IB method has recently also been used to impose solid boundary conditions in segregated multiphase flows simulations, where fluid/fluid interfaces are explicitly captured.
The majority of this past research has involved the combination of the volume-of-fluid multiphase model with the IB method in order to capture multiphase fluid/structure interaction.
The use of interface-capturing methods allows for the solid boundary to be accounted for using the same methods as single-phase IB methods, with the interface-capturing multiphase model account for fluid/fluid interfaces.
Applications of the IB method for interface-capturing include wave propagation \citep{Shen2008,Shen2010b,Shen2011c,Zhang2013,Zhang2014,Gsell2016,Yang2009}, injectors \citep{Suh2009}, porous media \citep{Patel2017}, hydroplaning \citep{Vincent2011} and capillary flow \citep{Horgue2014}.
Shen and Chan \citep{Shen2008,Shen2010b,Shen2011c}, Zhang and co-workers \citep{Zhang2013,Zhang2014}, \citet{Gsell2016} and \citet{Yang2009} independently coupled the IB method with an interface-capturing scheme to study the fluid-structure interaction of waves, validating with past experimental results.
\citet{Suh2009} developed a numerical method to model the piezoelectric inkjet process using IB with the level-set method with the droplet shape predicted by this method validated with analytical sharp-interface solutions for a range of contact angles.
\citet{Patel2017} used the IB method with the volume-of-fluid model to simulate water flooding processes encountered in enhanced oil recovery applications.
Capillary flow was captured using a similar approach by \citet{Horgue2014}, which was was validated using analytical solutions of pressure inside a droplet.

In addition to the IB method, multiphase fluid-solid interactions have been modeled using the fictitious domain and interface-capturing methods.
\citet{Vincent2011} modeled three-dimensional hydroplaning where the tire boundaries were captured using the fictitious domain method.
\citet{Arienti2014} combined the level-set and volume-of-fluid methods to model diesel injectors that showed good agreement with experimental results for predicting the mean axial velocity.
Similar to the IB studies, the models showed good agreement when validated against experimental results.
However, the use of interface-capturing methods with methods such as IB or fictitious domain severely limits the flow regimes that can be modeled since every fluid/fluid interface in the domain is resolved.

Interface-capturing multiphase models are infeasible for most chemical engineering processes, where dispersed multiphase flows are observed.
This multiphase flow regime involves a large surface area of fluid/fluid interfaces which are deformable, yielding the use of interface-capturing multiphase models infeasible.
Instead, the use of volume/time-averaged multiphase models, generally referred to as two-fluid models \citep{Ishii2011,Jakobsen2014} is required for simulations at experimentally and industrially relevant scales.
For example, bubble columns involve a liquid phase with large numbers of dispersed bubbles, where the presence of many evolving interfaces results in an infeasible computational cost for interface-capturing methods and relevant justification for the use of coarse-grained two-fluid models.

% nasser@tanya, I do not think we need to include this paragraph, I pulled pieces from this for the objective paragraph that follows
%The diffuse-interface method has been extensively used to model gas-liquid and liquid-liquid multiphase flows using the Cahn-Hilliard \citep{Abels2012, Abels2017, Ding2007, Jacqmin1999, Liu2015, Shen2010, Takada2006, Anderson1998} and Allen-Cahn \citep{Sun2007} equations.
%Additionally, the diffuse-interface method has been used to impose solid boundaries in simulations with interface-capturing \citep{Patel2018} and interface-resolution \citep{Aland2010}.
%This method does not require the solution to be interpolated at every time step since the phase-field ensures a smooth transition from the fluid to the solid phase.
%The boundary normal vector is computed from the gradient of the phase-field, aiding in the imposition of Neumann boundary conditions.
%Based on these factors, the diffuse-interface method is an attractive approach to imposing physical boundaries in simulations involving multiphase flow.

In this work, a novel diffuse solid-fluid interface method is presented for imposing solid boundaries in systems with dispersed multiphase flow conditions.
The diffuse-interface method has been extensively used to model gas-liquid and liquid-liquid multiphase flows using the Cahn-Hilliard \citep{Abels2012, Abels2017} and Allen-Cahn \citep{Sun2007} models.
The diffuse-interface method is applied in this work to capture physical boundaries, but now with a model for dispersed multiphase flows, specifically the two-fluid model \citep{Ishii2011}.
This approach allows for dispersed multiphase flow to be modeled without the need for re-meshing when the solid boundaries are evolved.
The method is presented and applied to model two-dimensional bubbly flow in a rectangular channel and bubbly flow with an immersed stationary cylinder and validated through comparison to simulation of the domains using the standard conformal mesh approach.

% tanya@nasser: I removed this since it no longer matches our outline
%The paper is organized as follows: background on the two-fluid model equations, presentation of the diffuse-interface method developed in this work and the numerical method used to solve the governing equations, simulation results on the effect of interface length-scale and function and conclusions.

\section{Results and Discussion}\label{sec:results}

To validate the use of the diffuse-interface method for imposing non-deformable solid boundaries, simulations of dispersed two-phase flow using the diffuse-interface are compared to simulation results from a boundary-conformal mesh for both channel flow and flow past a cylinder.
The effect of the diffuse-interface length-scale and function type on the solution and the performance of the method are discussed.

In this work, velocity fields are visualized using the line integral convolution (LIC) method \cite{Cabral1993,Laramee2003}, which enables significantly higher resolution of local flow alignment compared to streamlines along with the ability to superimpose coloring to indicate an additional scalar field (velocity magnitude, volume fraction, \textit{etc.}).

\subsection{Channel Flow}\label{sec:channel_flow}

The phase-field that defines the channel is described using the following hyperbolic tangent function:
\begin{equation}
    \phi(\tilde{\vb*{x}}) = \tanh(\frac{\abs{\tilde{x}} - \tilde{x}_c}{0.5\epsilon}),\label{eq:phi_channel_tanh}
\end{equation}
where $\tilde{x}_c=0.5$ is the scaled distance from the centerline to the channel wall and $\epsilon$ is a parameter associated with the width of the diffuse-interface.
The function asymptotically approaches $\phi=-1$ and $\phi=1$, resulting in a smooth transition between the phases, its value indicating each of the phases.
The scaled width of the interface, $\eta$, is approximated by the distance between $\phi=-0.999$ and $\phi=0.999$ which is given by $\eta=\epsilon\tanh[-1](0.999)$.
\nomenclature{$\eta$}{Diffuse-interface width}%

The presence of the diffuse-interface alters imposition of the no-slip boundary condition at the channel walls.
In the case of a boundary-conformal mesh, the velocities at the walls may be directly constrained to zero (stationary no-slip).
However, in the diffuse-interface method, the no-slip condition is blended with the governing equations for the two-fluid model.
The sharpness of the velocity gradient from the channel walls to the bulk is a function of the diffuse-interface function, interface width and discretization scheme.

In this study, all simulations use the same spatial discretization scheme and an embedded time-integration scheme in order to estimate the local error \citep{Ascher1998}.
Velocity gradients in the blended regions near the wall, resulting from the no-slip condition, are found to be the largest contributor to the local error which results in small time-steps required to impose the local error tolerance $\epsilon_l = \num{e-4}$.
This issue is particularly significant in cases where the diffuse-interface is large such as in channel flow.
To mitigate this constraint on the time-step size, only the local error inside the fluid domain, where $\phi \leq -0.999$, is considered when computing the new step size and the local error tolerance is relaxed to $\epsilon_l = \num{e-3}$.

\begin{figure}
    \centering
    \begin{subfigure}{0.45\linewidth}
        \includegraphics[width=\textwidth]{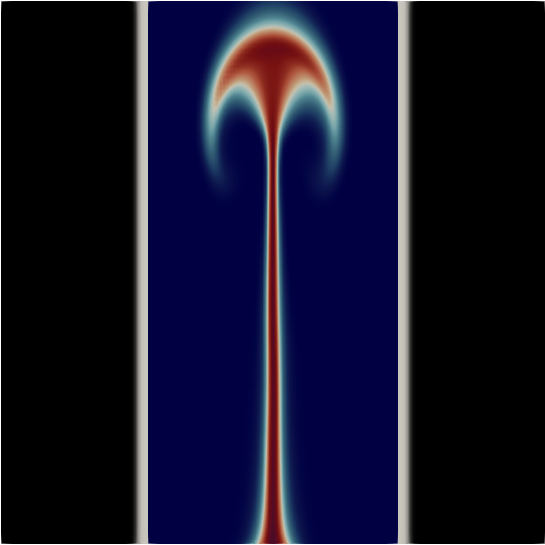}
        \caption*{}
    \end{subfigure}
    \begin{subfigure}{0.078\linewidth}
        \includegraphics[width=\linewidth]{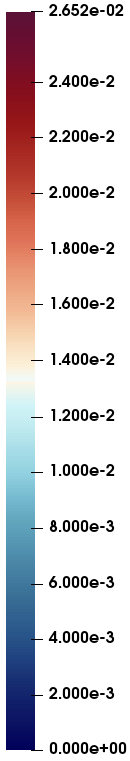}
        \caption*{$\alpha_{g}$}
    \end{subfigure}
    \begin{subfigure}{0.055\linewidth}
        \includegraphics[width=\linewidth]{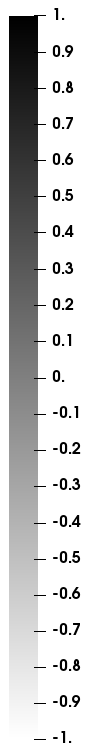}
        \caption*{$\phi$}
    \end{subfigure}
    \caption{Surface plot of $\alpha_g$ at $t=\SI{1.72}{\second}$ with hyperbolic tangent diffuse-interface and $\epsilon = 0.02$.
    The $\phi$ profile is superimposed and thresholded show only $\phi \geq -0.999$.
    The gray-scale color bar denotes the phase-field that describes the diffuse-interface, thresholded to show $\phi \geq -0.999$.}
    \label{fig:tanh01_alpha_g_phi}
\end{figure}

\begin{figure}
    \centering
    \begin{subfigure}{0.25\linewidth}
        \includegraphics[width=\linewidth]{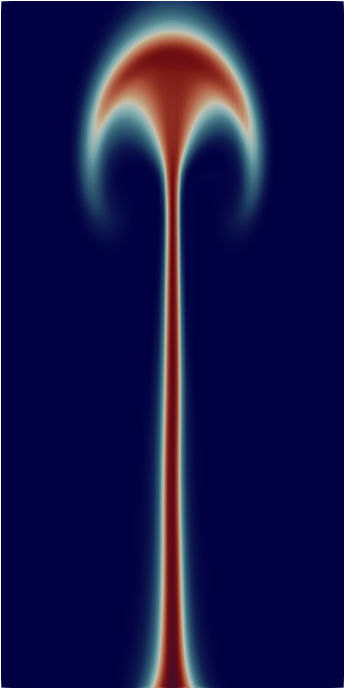}
        \caption*{$\alpha_g$}\label{fig:var_CD_bounded_intp}
    \end{subfigure}
    \begin{subfigure}{0.25\linewidth}
        \includegraphics[width=\linewidth]{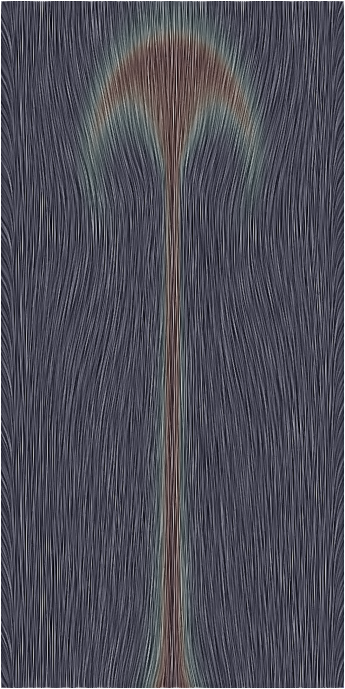}
        \caption*{$\vb*{v}_g$}\label{fig:var_CD_bounded_intp_vg}
    \end{subfigure}
    \begin{subfigure}{0.25\linewidth}
        \includegraphics[width=\linewidth]{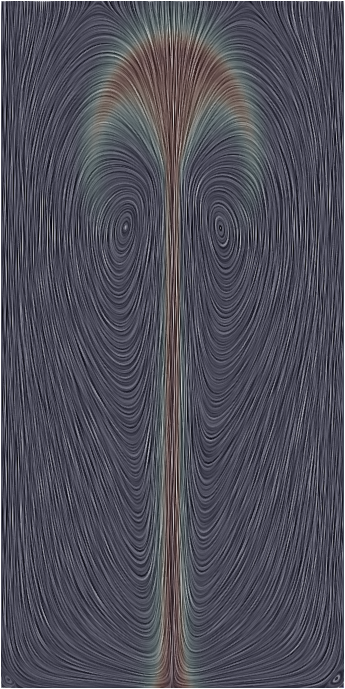}
        \caption*{$\vb*{v}_l$}\label{fig:var_CD_bounded_intp_vl}
    \end{subfigure}
    \begin{subfigure}{0.095\linewidth}
        \includegraphics[width=\linewidth]{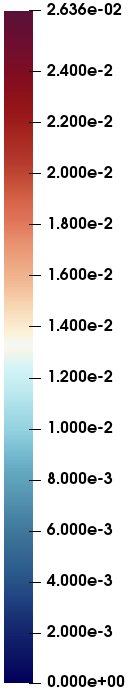}
        \caption*{}
    \end{subfigure}
    \caption{Surface plot of (left) phase fraction, (center) gas velocity and (right) liquid velocity at $t=\SI{1.72}{\second}$ from bounded IPCS with interfacial pressure. Reprinted from \citet{Treeratanaphitak2019} with permission from Elsevier.}
    \label{fig:var_CD_intp}
\end{figure}

\begin{figure}
    \centering
    \begin{subfigure}{0.25\linewidth}
        \includegraphics[width=\linewidth]{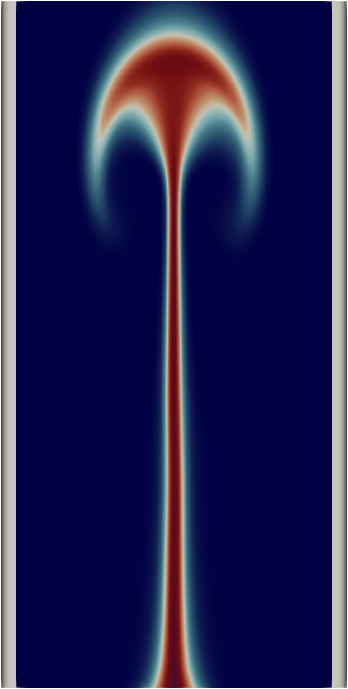}
        \caption*{$\alpha_g$}\label{fig:tanh01_alpha_g}
    \end{subfigure}
    \begin{subfigure}{0.251\linewidth}
        \includegraphics[width=\linewidth]{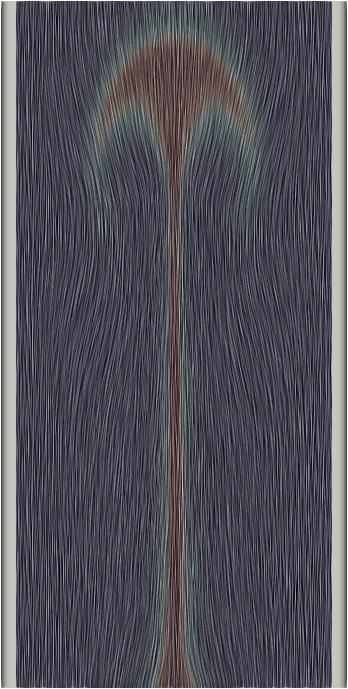}
        \caption*{$\vb*{v}_g$}\label{fig:tanh01_v_g}
    \end{subfigure}
    \begin{subfigure}{0.251\linewidth}
        \includegraphics[width=\linewidth]{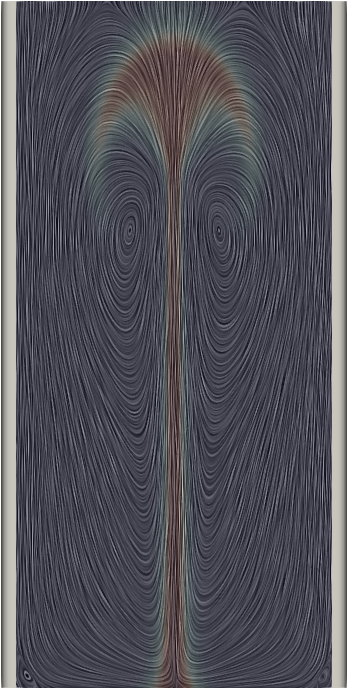}
        \caption*{$\vb*{v}_l$}\label{fig:tanh01_v_l}
    \end{subfigure}
    \begin{subfigure}{0.086\linewidth}
        \includegraphics[width=\linewidth]{TFM_DI/channel/tanh01/alpha_g_cbar}
        \caption*{}
    \end{subfigure}
    \begin{subfigure}{0.06\linewidth}
        \includegraphics[width=\linewidth]{TFM_DI/phi_cbar}
        \caption*{$\phi$}
    \end{subfigure}
    \caption{Surface plot of (left) $\alpha_g$, (center) gas velocity and (right) liquid velocity at $t=\SI{1.72}{\second}$ with hyperbolic tangent diffuse-interface and $\epsilon = 0.02$.}
    \label{fig:tanh01}
\end{figure}

The gas phase fraction profile at $t=\SI{1.72}{\second}$ obtained from simulation with a diffuse-interface given by \cref{eq:phi_channel_tanh} and $\epsilon = 0.02$ is shown in \cref{fig:tanh01_alpha_g_phi}.
Qualitatively, the phase fraction profile and transient behavior are in agreement with that observed in past work \citep{Treeratanaphitak2019}, where traditional conformal mesh simulations are carried out using the multiphase finite element-based solver also used in this work.
A bubble plume is formed as the dispersed gas phase flows through the liquid phase, where the unidirectional flow of gas phase imparts recirculatory flow of the liquid phase.
Over time, the plume increases in width, driven by the dispersive action of vortices formed in the wake of the plume.
This is in qualitative agreement with experimental observations of the startup period in rectangular bubble columns \citep{Mudde2005}.
\Cref{fig:tanh01} shows the gas and liquid velocity LICs inside the box given by $x\in [-0.025,0.025]$ and $y\in[0,0.1]$ at the same simulation time step.
From \cref{fig:tanh01,fig:var_CD_intp}, the velocity profiles of both gas and liquid phases are similar with liquid recirculating in the wake of the bubble plume.

In addition to qualitative comparisons of the phase fraction profile and velocity LICs, the time evolution of the gas hold-up from the diffuse-interface simulation is also be compared to that of the reference solution from \citet{Treeratanaphitak2019}.
The gas hold-up in the diffuse-interface simulation is determined as follows:
\begin{equation}
    \langle\alpha_g\rangle = \frac{\int_{\Omega} \frac{1-\phi}{2} \alpha_g\dd\Omega}{\int_{\Omega}\frac{1-\phi}{2}\dd\Omega},\label{eq:avg_holdup_phi}
\end{equation}
where the denominator is the volume of the physical domain.
This comparison is reported in the following sections.

\subsubsection{Effect of Interface Length-Scale} \label{sec:effect_length_scale}

\begin{figure}
    \centering
    \includegraphics[width=0.75\textwidth]{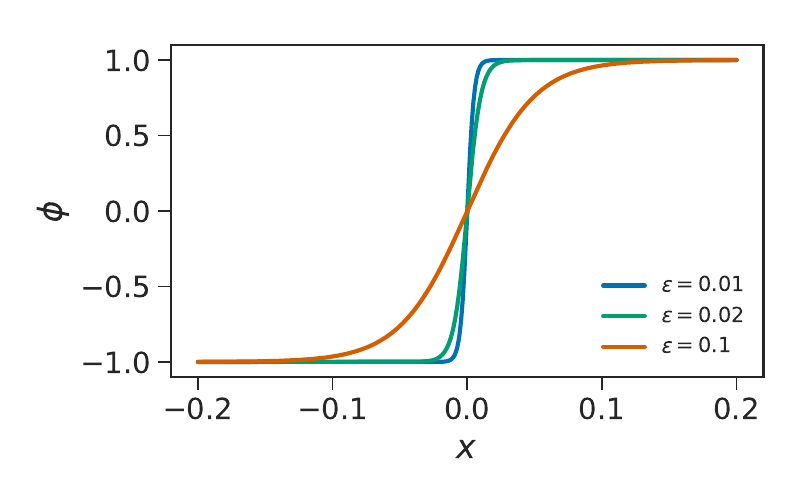}
    \caption{Comparison of diffuse-interface width generated using the the hyperbolic tangent function with varying $\epsilon$.}
    \label{fig:phi_tanh_ep_comp}
\end{figure}

The blending of the conservation equations and boundary conditions of the solid and multiphase fluid resulting from the introduction of the diffuse-interface may affect the accuracy of simulation results, compared to reference boundary-conformal mesh solutions.
In this section, a study is performed to determine the effect of the diffuse interface length-scale on the accuracy.
Simulations of the channel flow with the same geometry as before are repeated for a range of diffuse interface widths, $\epsilon = ~$\numlist{0.01;0.02;0.04;0.08;0.1}.
\Cref{fig:phi_tanh_ep_comp} shows how the $\phi = \tanh(x/0.5\epsilon)$ profile changes with different values of $\epsilon$.
A sharper (less) diffuse interface corresponds to $\epsilon = 0.01$ with a wider (more) diffuse interface corresponding to $\epsilon = 0.1$, for example.

\begin{figure}
    \centering
    \begin{subfigure}{0.45\linewidth}
        \includegraphics[width=\textwidth]{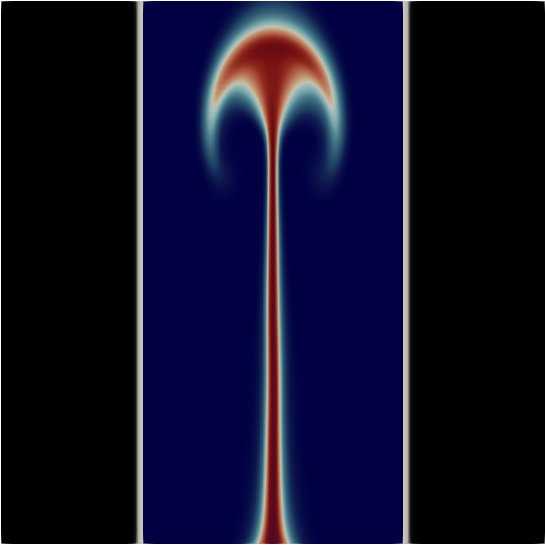}
        \caption*{}
    \end{subfigure}
    \begin{subfigure}{0.078\linewidth}
        \includegraphics[width=\linewidth]{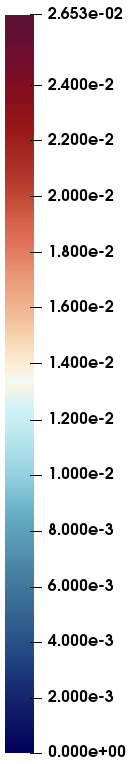}
        \caption*{$\alpha_g$}
    \end{subfigure}
    \begin{subfigure}{0.055\linewidth}
        \includegraphics[width=\linewidth]{TFM_DI/phi_cbar}
        \caption*{$\phi$}
    \end{subfigure}
    \caption{Surface plot of $\alpha_g$ at $t=\SI{1.72}{\second}$ with hyperbolic tangent diffuse-interface and $\epsilon = 0.01$.
    The gray-scale color bar denotes the phase-field that describes the diffuse-interface, thresholded to show $\phi \geq -0.999$.}
    \label{fig:tanh005_alpha_g_phi}
\end{figure}

\begin{figure}
    \centering
    \begin{subfigure}{0.45\linewidth}
        \includegraphics[width=\textwidth]{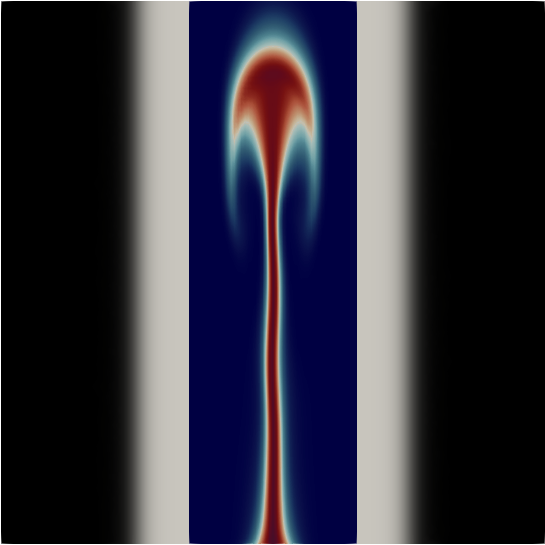}
        \caption*{}
    \end{subfigure}
    \begin{subfigure}{0.078\linewidth}
        \includegraphics[width=\linewidth]{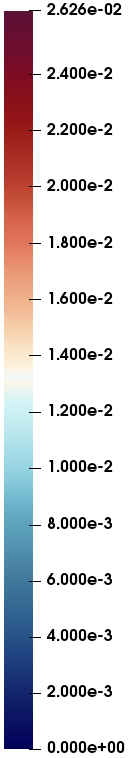}
        \caption*{$\alpha_g$}
    \end{subfigure}
    \begin{subfigure}{0.055\linewidth}
        \includegraphics[width=\linewidth]{TFM_DI/phi_cbar}
        \caption*{$\phi$}
    \end{subfigure}
    \caption{Surface plot of $\alpha_g$ at $t=\SI{1.72}{\second}$ with hyperbolic tangent diffuse-interface and $\epsilon = 0.1$. 
    The gray-scale color bar denotes the phase-field that describes the diffuse-interface, thresholded to show $\phi \geq -0.999$.}
    \label{fig:tanh05_alpha_g_phi}
\end{figure}

As the diffuse interface width increases and the interface becomes more diffuse, the contribution of local error from blending increases.
However, given that this contribution to the local error at every time-step is spatially localized to the blending region, the time-step size is comparable between all values of $\epsilon$ through the use of the local error approach presented in Section \ref{sec:channel_flow}.
\Cref{fig:tanh005_alpha_g_phi,fig:tanh05_alpha_g_phi} show the gas phase fraction at $t=\SI{1.72}{\second}$ for simulations with $\epsilon=0.01$ and $\epsilon=0.1$, respectively.
Qualitatively, the gas fraction profile from $\epsilon=0.01$ is nearly identical to the case with $\epsilon=0.02$, but the profile from $\epsilon=0.1$ is notably different from $\epsilon=0.02$.
In \cref{fig:tanh05_alpha_g_phi}, there is a noticeable modulation of the gas column below the plume and the plume is much narrower.
This is due to the interface being very diffuse and the effect of the solid boundary conditions is blended further into the fluid domain.

\begin{figure}
    \centering
    \begin{subfigure}{0.25\linewidth}
        \includegraphics[width=\linewidth]{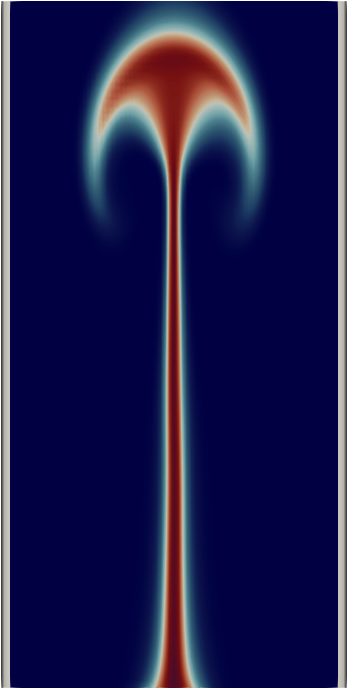}
        \caption*{$\alpha_g$}\label{fig:tanh005_alpha_g}
    \end{subfigure}
    \begin{subfigure}{0.251\linewidth}
        \includegraphics[width=\linewidth]{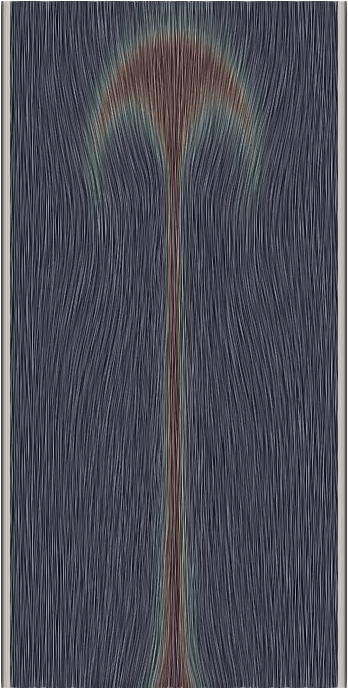}
        \caption*{$\vb*{v}_g$}\label{fig:tanh005_v_g}
    \end{subfigure}
    \begin{subfigure}{0.251\linewidth}
        \includegraphics[width=\linewidth]{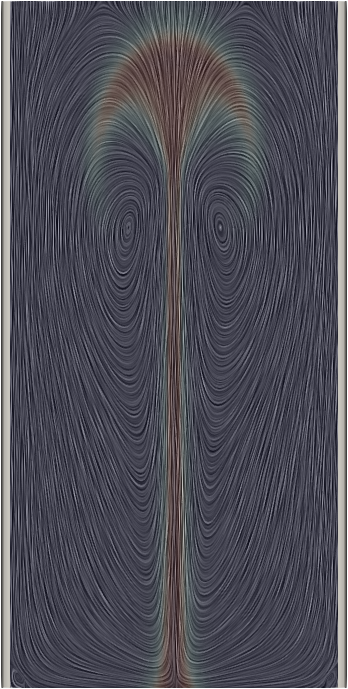}
        \caption*{$\vb*{v}_l$}\label{fig:tanh005_v_l}
    \end{subfigure}
    \begin{subfigure}{0.086\linewidth}
        \includegraphics[width=\linewidth]{TFM_DI/channel/tanh005/alpha_g_cbar}
        \caption*{}
    \end{subfigure}
    \begin{subfigure}{0.06\linewidth}
        \includegraphics[width=\linewidth]{TFM_DI/phi_cbar}
        \caption*{$\phi$}
    \end{subfigure}
    \caption{Surface plot of (left) $\alpha_g$, (center) gas velocity and (right) liquid velocity at $t=\SI{1.72}{\second}$ with hyperbolic tangent diffuse-interface and $\epsilon = 0.01$.}
    \label{fig:tanh005}
\end{figure}

\begin{figure}
    \centering
    \begin{subfigure}{0.25\linewidth}
        \includegraphics[width=\linewidth]{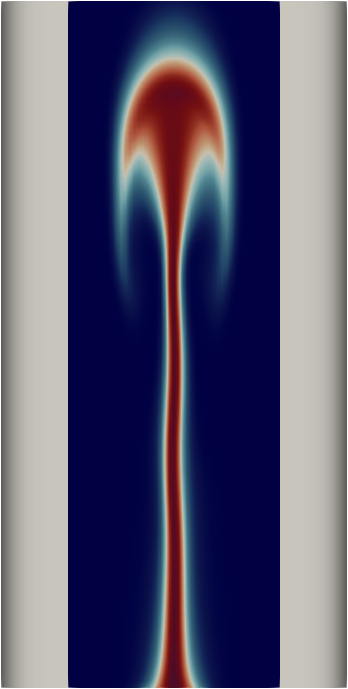}
        \caption*{$\alpha_g$}\label{fig:tanh05_alpha_g}
    \end{subfigure}
    \begin{subfigure}{0.251\linewidth}
        \includegraphics[width=\linewidth]{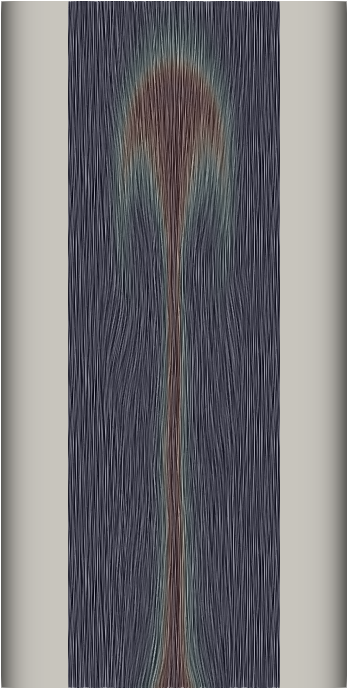}
        \caption*{$\vb*{v}_g$}\label{fig:tanh05_v_g}
    \end{subfigure}
    \begin{subfigure}{0.251\linewidth}
        \includegraphics[width=\linewidth]{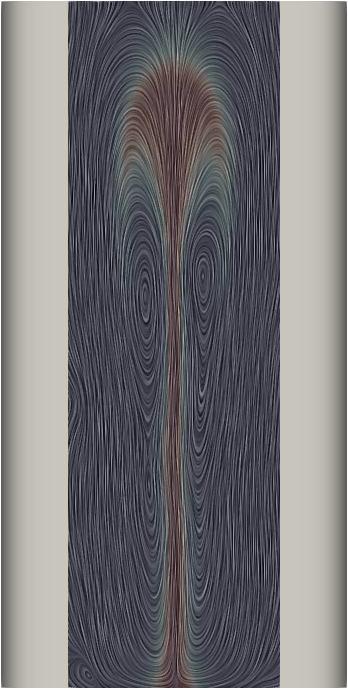}
        \caption*{$\vb*{v}_l$}\label{fig:tanh05_v_l}
    \end{subfigure}
    \begin{subfigure}{0.086\linewidth}
        \includegraphics[width=\linewidth]{TFM_DI/channel/tanh05/alpha_g_cbar}
        \caption*{}
    \end{subfigure}
    \begin{subfigure}{0.06\linewidth}
        \includegraphics[width=\linewidth]{TFM_DI/phi_cbar}
        \caption*{$\phi$}
    \end{subfigure}
    \caption{Surface plot of (left) $\alpha_g$, (center) gas velocity and (right) liquid velocity at $t=\SI{1.72}{\second}$ with hyperbolic tangent diffuse-interface and $\epsilon = 0.1$.}
    \label{fig:tanh05}
\end{figure}

The gas and liquid velocity LICs from $\epsilon=0.01$ and $\epsilon=0.1$ are shown in \cref{fig:tanh005,fig:tanh05}, respectively.
The LICs from $\epsilon=0.01$ are qualitatively similar to those observed in \cref{fig:var_CD_intp,fig:tanh01}.
However, the LICs from $\epsilon=0.1$ are different from the other simulations.
The gas velocity LICs appear to exhibit less curvature in the wake of the bubble plume and the liquid velocity vortices in the wake of the plume are narrower due to the highly diffuse nature of the interface.

\begin{figure}
    \centering
    \includegraphics[width=0.75\linewidth]{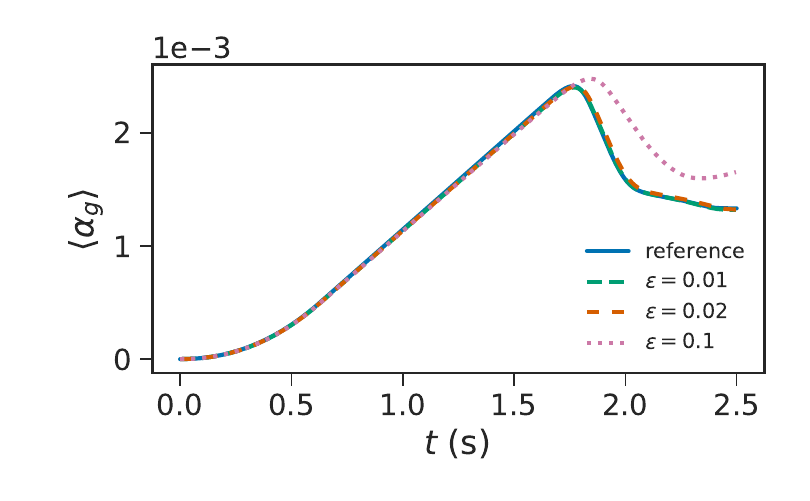}
    \caption{Time evolution of overall gas hold-up inside a channel with solid boundaries defined by a hyperbolic tangent diffuse-interface.}\label{fig:int_alpha_g_comp_channel_tanh}
\end{figure}

\Cref{fig:int_alpha_g_comp_channel_tanh} shows the time evolution of the overall gas hold-up, $\langle\alpha_g\rangle$, inside the channel up to $\SI{2.5}{\second}$ from the diffuse interface (hyperbolic tangent variation) simulations and the reference (conformal mesh) solution.
At narrow interface widths, the evolution of the gas hold-up follows the same evolution as the reference solution and the magnitude of the overall hold-up is equivalent.
However, for $\epsilon=0.1$, the evolution of the hold-up is similar to the reference solution only up to the point where the bubble plume leaves the channel.
After this point, the gas hold-up deviates from the reference solution, indicating that the flow behavior is different.
In the reference solution, after the transient convection of the bubble plume, a straight vertical column of bubbly flow is observed.
In the case of $\epsilon=0.1$, the column of bubbly flow undulates (\cref{fig:tanh01}) and the onset of vertical column flow occurs much earlier than in the other simulations.

\begin{figure}
    \centering
    \begin{subfigure}{0.45\textwidth}
        \centering
        \includegraphics[width=\textwidth]{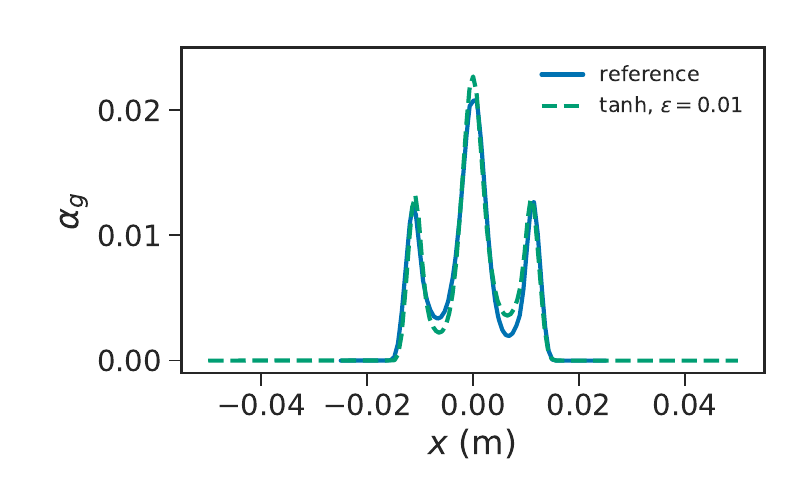}
        \caption{$\epsilon = 0.01$}\label{fig:tanh_ref_comp005}
    \end{subfigure}
    \begin{subfigure}{0.45\textwidth}
        \centering
        \includegraphics[width=\textwidth]{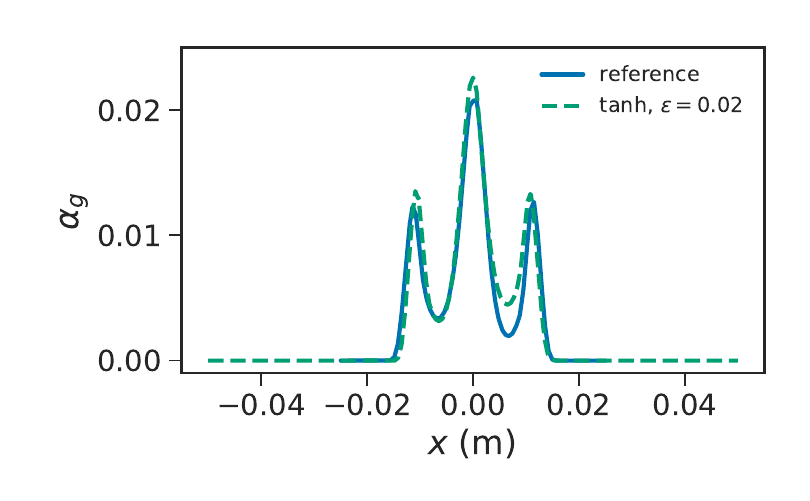}
        \caption{$\epsilon = 0.02$}\label{fig:tanh_ref_comp01}
    \end{subfigure}
    \\
    \begin{subfigure}{0.45\textwidth}
        \centering
        \includegraphics[width=\textwidth]{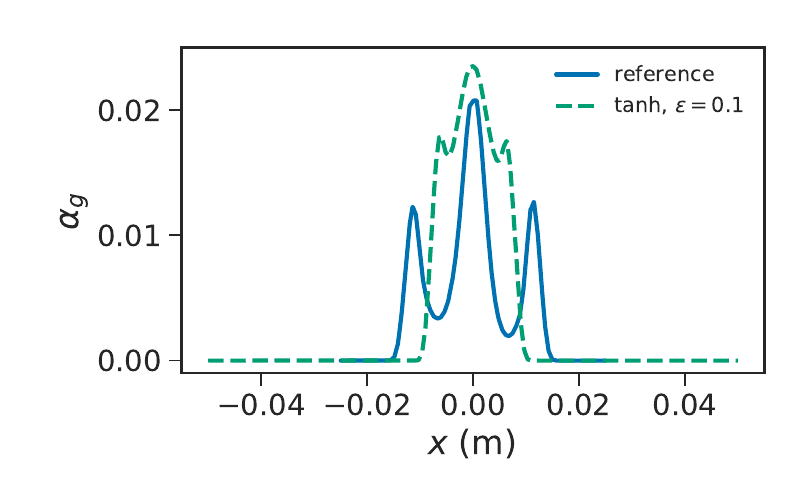}
        \caption{$\epsilon = 0.1$}\label{fig:tanh_ref_comp05}
    \end{subfigure}
    \caption{$\alpha_g$ profile along $y=\SI{0.08}{\meter}$ with different hyperbolic tangent diffuse-interface widths.}\label{fig:tanh_ref_comp}
\end{figure}

The gas fraction is sampled along the line $y=\SI{0.08}{\meter}$, which corresponds to the widest part of the bubble plume, and the profile along the $x$-axis is plotted in \cref{fig:tanh_ref_comp}.
For the cases where $\epsilon = 0.01$ and $\epsilon = 0.02$, the $\alpha_g$ profiles obtained using a diffuse-interface method show good qualitative agreement with the reference solution from \citet{Treeratanaphitak2019}.
This qualitative agreement improves as the diffuse-interface is reduced, with simulation results being almost equivalent for the smallest diffuse interface.
However, as the interface becomes wider the $\alpha_g$ profile deviates from the reference solution, which is intuitive.
The effect of the wide diffuse-interface is clear as $\alpha_g$, shown in \cref{fig:tanh_ref_comp05}, starts to transition from $\alpha_g = 0$ to a nonzero value further into the domain.

To obtain a quantitative measure of the difference between simulation results using the diffuse-interface  versus the reference solution, the width of the bubble plume at $y=\SI{0.08}{\meter}$ is computed and shown in \cref{tab:plume_width_tanh}.
The plume width from simulations with $\epsilon=0.01$ and $\epsilon=0.02$ are within $3\%$ of the reference solution, supporting the accuracy observations mentioned earlier.
The plume width from the simulation with $\epsilon=0.1$ differs by $30\%$ from that of reference solution, highlighting the importance of appropriate choice of the diffuse-interface width.

\begin{table}
    \centering
    \caption{Bubble plume width at $y = \SI{0.08}{\meter}$ from simulations using hyperbolic tangent diffuse-interface.}
    \label{tab:plume_width_tanh}
    \begin{tabular}{ccc}
        \hline
        Study & $x_{\text{plume}}$ ($\times \num{e-2}~\si{\meter}$) & Difference (\%) \\
        \hline
        Reference & $3.21$ & -- \\
        $\epsilon = 0.01$ & $3.17$ & $1.25$ \\
        $\epsilon = 0.02$ & $3.12$ & $2.80$ \\
        $\epsilon = 0.04$ & $2.92$ & $9.03$ \\
        $\epsilon = 0.08$ & $2.51$ & $21.8$ \\
        $\epsilon = 0.1$ & $2.25$ & $30.0$ \\
        \hline
    \end{tabular}
\end{table}

\subsubsection{Effect of Interface Function} \label{sec:effect_interface_fn}

The usage of the hyperbolic tangent function as the kernel for the diffuse-interface is generally the most common approach \citep{Nguyen2018}, but other functions have been used that result in a continuous transition from the indicator values for the solid to the fluid regions \citep{Abels2012}.
An example of an alternate kernel function is piece-wise cosine where the interface region is described by a cosine function that is between $[-1,1]$ and outside the interface region, $\phi = \pm 1$.
Unlike the hyperbolic tangent function, which asymptotically approaches the lower and upper bounds of $\phi$, the piece-wise cosine function reaches these values $\phi = \pm 1$ exactly at the specified $\eta$.
In this section, the effects of using the following piece-wise cosine function in the presented diffuse interface method is studied:
\begin{equation}
\phi(\tilde{\vb*{x}}) = -\cos(-\pi\min\qty[1, \max\qty(0,\frac{\abs{\tilde{x}}-\tilde{x}_c+0.5\eta}{\eta})]),\label{eq:phi_channel_cos}
\end{equation}
where $\phi$ will be $\pm 1$ outside the region $\tilde{x} \in (\tilde{x}_c - 0.5\eta, \tilde{x}_c + 0.5\eta)$, depending on which side of the channel wall $\tilde{x}$ is close to.

\begin{figure}
    \centering
    \includegraphics[width=0.75\textwidth]{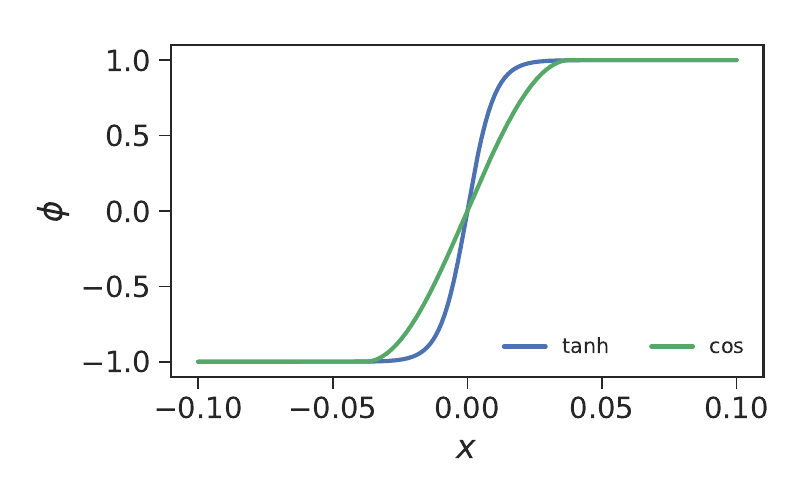}
    \caption{Comparison of diffuse-interface generated using hyperbolic tangent and piece-wise cosine functions with $\epsilon = 0.02$ and $\eta = \epsilon\tanh[-1](0.999)$.}
    \label{fig:tanh_vs_cos}
\end{figure}

\Cref{fig:tanh_vs_cos} shows the phase field $\phi$ profile variation with respect to $x$ when defined using a hyperbolic tangent function, $\phi = \tanh(x/0.5\epsilon)$ and using a piece-wise cosine function centered at $x_c=0$, $\phi = -\cos(-\pi\min\qty[1, \max\qty(0,(x+0.5\eta)/\eta)])$, for a comparable interface width.
The width of the cosine interface is approximated by $\eta = \epsilon\tanh[-1](0.999)$, which corresponds to the distance between $\phi = \pm 0.999$ in the hyperbolic tangent case.
From \cref{fig:tanh_vs_cos}, the transition of $\phi$ from $-1$ to $1$ in the piece-wise cosine function is more gradual than the hyperbolic tangent function, which results in lower values of $\grad{\phi}$.

\begin{figure}
    \centering
    \begin{subfigure}{0.2\linewidth}
        \includegraphics[width=\linewidth]{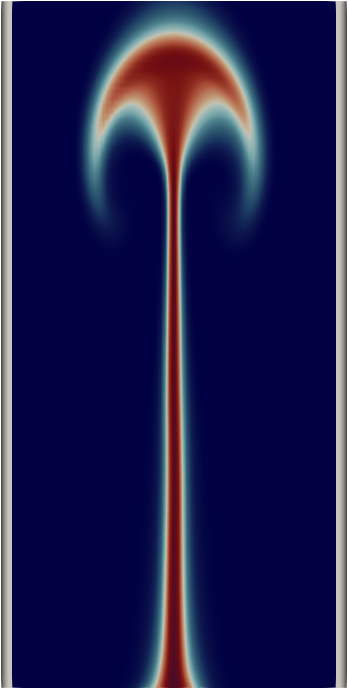}
    \end{subfigure}
    \begin{subfigure}{0.2\linewidth}
        \includegraphics[width=\linewidth]{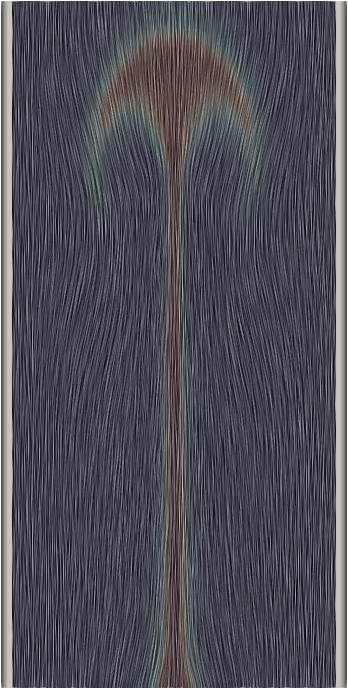}
    \end{subfigure}
    \begin{subfigure}{0.2\linewidth}
        \includegraphics[width=\linewidth]{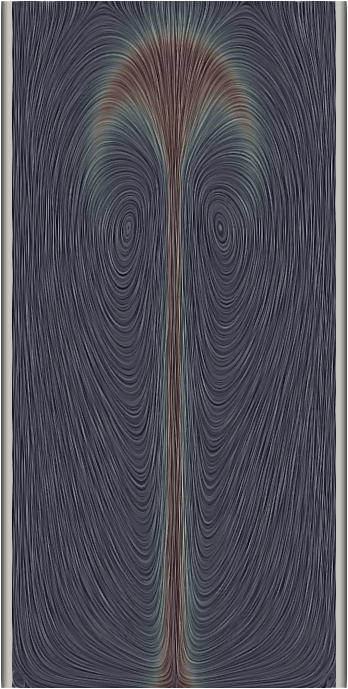}
    \end{subfigure}
    \begin{subfigure}{0.07\linewidth}
        \includegraphics[width=\linewidth]{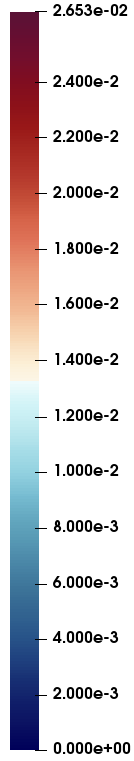}
    \end{subfigure}
    \begin{subfigure}{0.0475\linewidth}
        \includegraphics[width=\linewidth]{TFM_DI/phi_cbar}
    \end{subfigure}
    \\
    \begin{subfigure}{0.2\linewidth}
        \includegraphics[width=\linewidth]{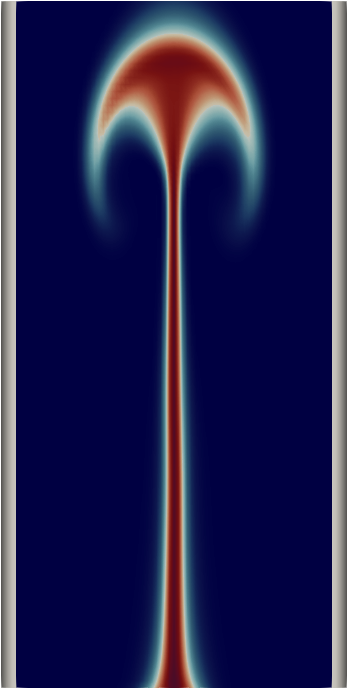}
    \end{subfigure}
    \begin{subfigure}{0.2\linewidth}
        \includegraphics[width=\linewidth]{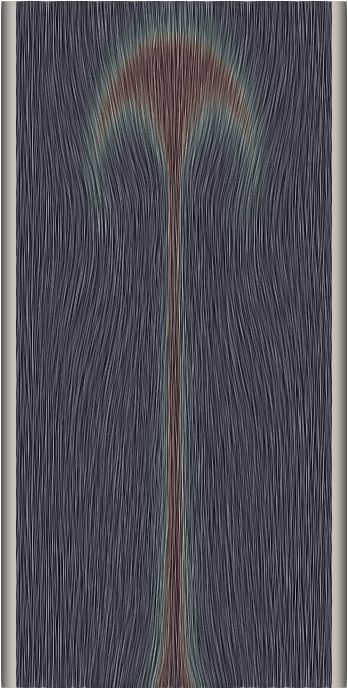}
    \end{subfigure}
    \begin{subfigure}{0.2\linewidth}
        \includegraphics[width=\linewidth]{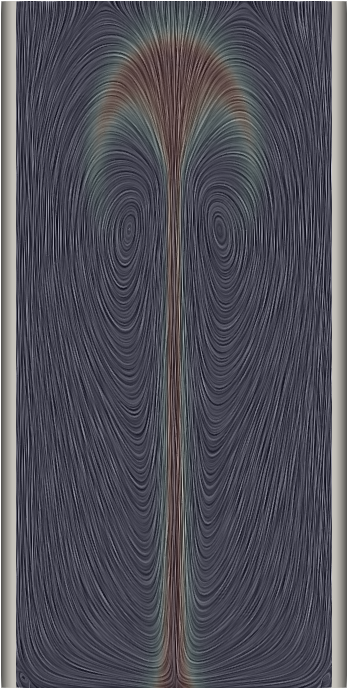}
    \end{subfigure}
    \begin{subfigure}{0.068\linewidth}
        \includegraphics[width=\linewidth]{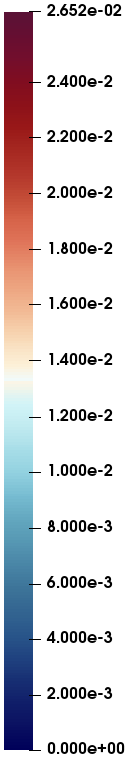}
    \end{subfigure}
    \begin{subfigure}{0.0475\linewidth}
        \includegraphics[width=\linewidth]{TFM_DI/phi_cbar}
    \end{subfigure}
    \\
    \begin{subfigure}{0.2\linewidth}
        \includegraphics[width=\linewidth]{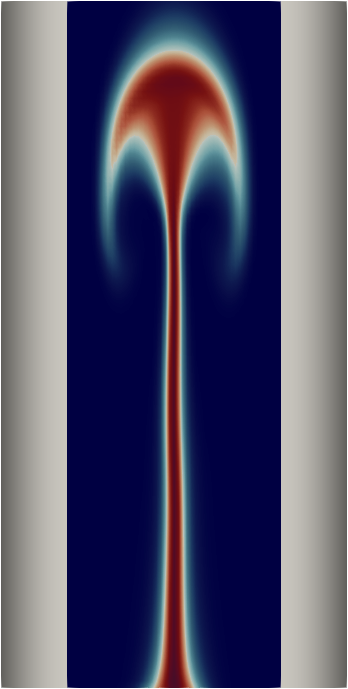}
        \caption*{$\alpha_g$}
    \end{subfigure}
    \begin{subfigure}{0.2\linewidth}
        \includegraphics[width=\linewidth]{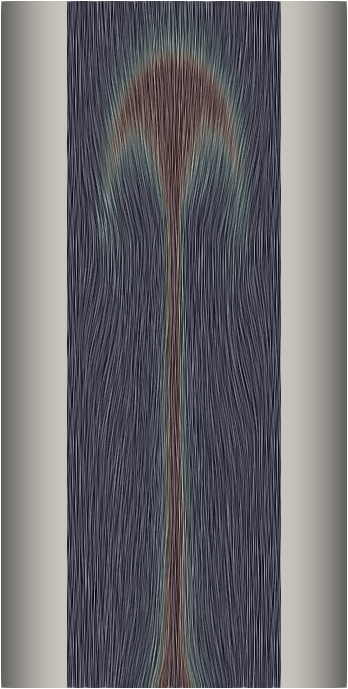}
        \caption*{$\vb*{v}_g$}
    \end{subfigure}
    \begin{subfigure}{0.2\linewidth}
        \includegraphics[width=\linewidth]{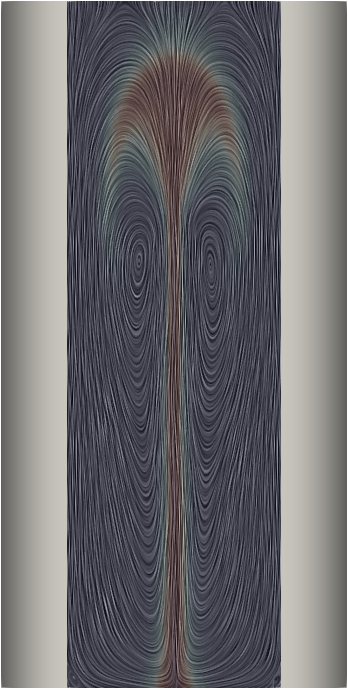}
        \caption*{$\vb*{v}_l$}
    \end{subfigure}
    \begin{subfigure}{0.068\linewidth}
        \includegraphics[width=\linewidth]{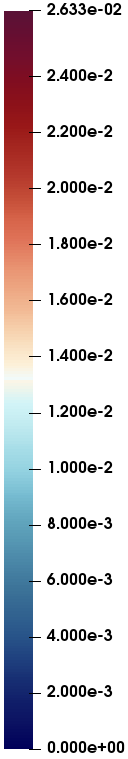}
        \caption*{}
    \end{subfigure}
    \begin{subfigure}{0.0475\linewidth}
        \includegraphics[width=\linewidth]{TFM_DI/phi_cbar}
        \caption*{$\phi$}
    \end{subfigure}
    \caption{Surface plot of (left) phase fraction, (center) gas velocity and (right) liquid velocity at $t=\SI{1.72}{\second}$ with piece-wise cosine diffuse-interface and (top) $\epsilon = 0.01$, (middle) $\epsilon = 0.02$ and (bottom) $\epsilon = 0.1$.}
    \label{fig:channel_cos}
\end{figure}

\Cref{fig:channel_cos} shows the gas fraction profile and velocity LICs for simulations with a piece-wise cosine diffuse-interface with a comparable interface width as the hyperbolic tangent case.
At small diffuse interface width $\epsilon$, the resulting simulation results are qualitatively similar to those results using the hyperbolic tangent.
The bubble plume in the $\epsilon=0.1$ simulation case is significantly narrower than the reference conformal mesh solution, but is wider than the corresponding hyperbolic tangent simulation result.
The column-like flow of gas plume also appears to be more stable than the results in \cref{fig:tanh05}.

The significant difference between the results from different interface functions at $\epsilon=0.1$ is attributed to the lack of asymptotic behavior of the piece-wise cosine, shown in \cref{eq:phi_channel_cos}, compared to that of the hyperbolic tangent function.
For the piece-wise cosine function, the approximation $\eta = \epsilon\tanh[-1](0.999)$ results in a diffuse-interface that approaches $\phi = \pm 1$ over a similar length-scale as the hyperbolic tangent function for much smaller interface widths.
However, at $\epsilon=0.1$, the difference between $\epsilon\tanh[-1](0.999)$ and $\epsilon\tanh[-1](0.9999)$, which are interface widths approximated by $\phi = \pm 0.999$ and $\phi = \pm 0.9999$, respectively, is an order of magnitude larger than at $\epsilon=0.01$ and non-negligible.
The hyperbolic tangent function diffuses the interface over a larger distance which, for larger values of $\epsilon$, is detrimental to the performance of the method.

\begin{figure}
    \centering
    \includegraphics[width=0.75\linewidth]{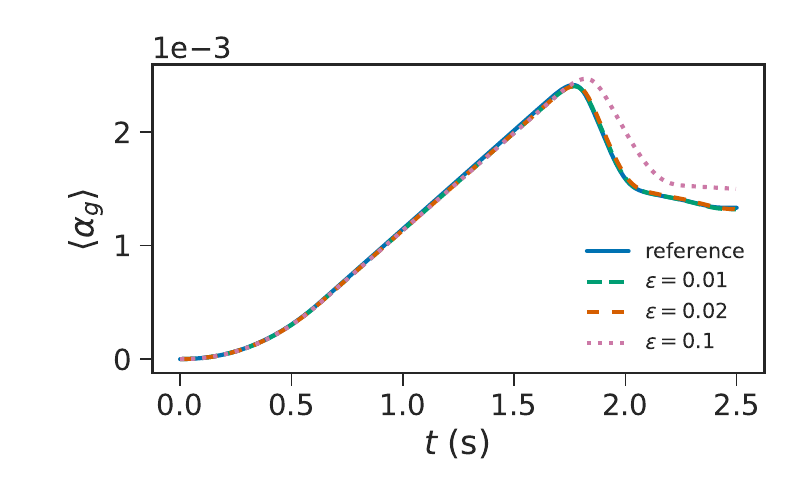}
    \caption{Time evolution of overall gas hold-up inside a channel with solid boundaries defined by a piece-wise cosine diffuse-interface.}\label{fig:int_alpha_g_comp_channel_cos}
\end{figure}

\Cref{fig:int_alpha_g_comp_channel_cos} shows the time evolution of the overall gas holdup for simulations with a piece-wise cosine diffuse-interface.
Similar to the hyperbolic tangent case, the gas hold-up at small interface widths ($\epsilon=0.01$ and $\epsilon=0.02$) are in agreement with the reference solution.
For the $\epsilon=0.1$ case, the gas hold-up differs from the reference solution as the bubble plume exits the simulation domain, but the difference is not as significant as the hyperbolic tangent case in \cref{fig:int_alpha_g_comp_channel_tanh}.

\begin{figure}
    \centering
    \begin{subfigure}{0.45\textwidth}
        \centering
        \includegraphics[width=\textwidth]{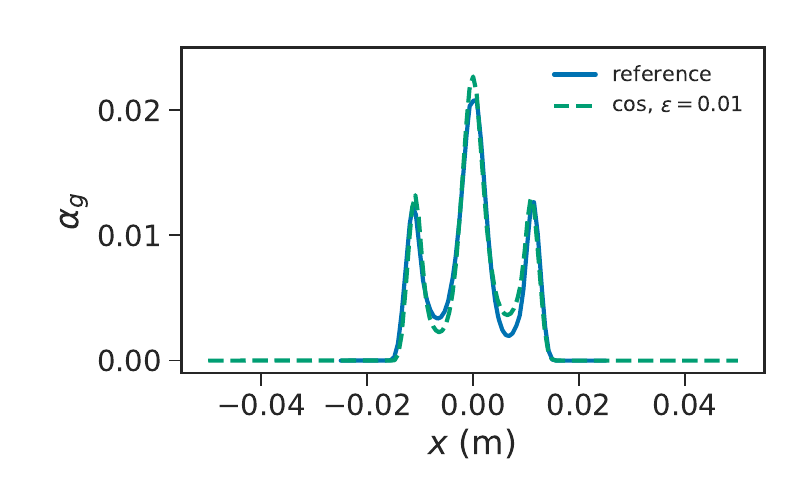}
        \caption{$\epsilon = 0.01$}\label{fig:cos_ref_comp005}
    \end{subfigure}
    \begin{subfigure}{0.45\textwidth}
        \centering
        \includegraphics[width=\textwidth]{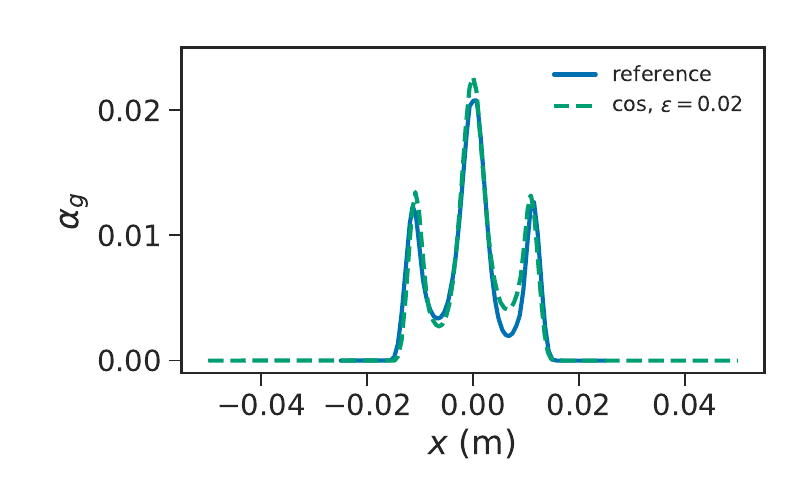}
        \caption{$\epsilon = 0.02$}\label{fig:cos_ref_comp01}
    \end{subfigure}
    \\
    \begin{subfigure}{0.45\textwidth}
        \centering
        \includegraphics[width=\textwidth]{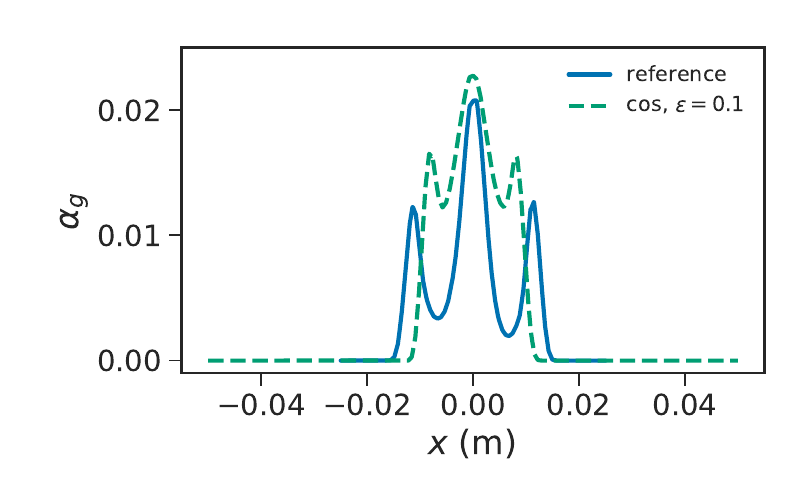}
        \caption{$\epsilon = 0.1$}\label{fig:cos_ref_comp05}
    \end{subfigure}
    \caption{$\alpha_g$ profile along $y=\SI{0.08}{\meter}$ with different piece-wise cosine diffuse-interface widths.}\label{fig:cos_ref_comp}
\end{figure}

\begin{figure}
    \centering
    \includegraphics[width=0.75\linewidth]{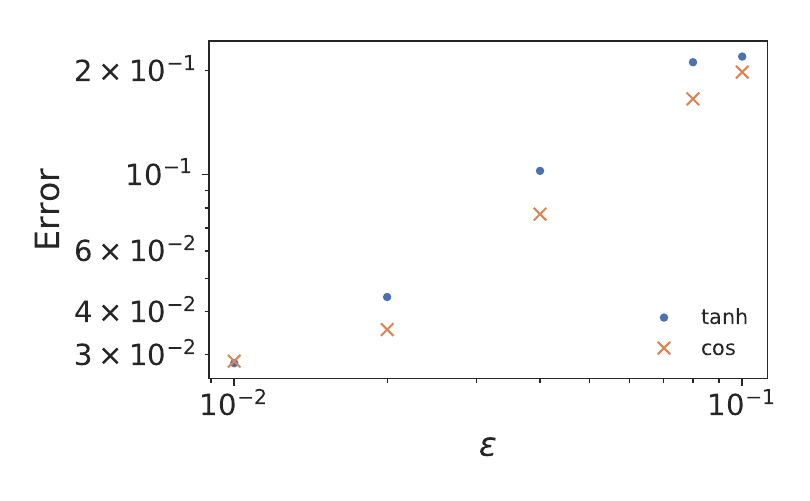}
    \caption{Error in $\alpha_g$ profile along $y=\SI{0.08}{\meter}$ as a function of $\epsilon$.}\label{fig:channel_error_all}
\end{figure}

The $\alpha_g$ profile at $y=\SI{0.08}{\meter}$ from the three cases are plotted with the reference solution in \cref{fig:cos_ref_comp}.
The results are similar to those observed in the previous section where simulations with $\epsilon=0.01$ and $\epsilon=0.02$ yield profiles that are comparable to the reference solution, but the profile from the simulation with $\epsilon=0.1$ yields a significantly different solution.
\Cref{fig:channel_error_all} describes the error in the phase fraction along the line $y=\SI{0.08}{\meter}$ where the interface width is varied for both interface functions.
The error is defined as:
\begin{equation}
\text{Error}=\norm{\alpha_{g,ref} - \alpha_g}_{y=\SI{0.08}{\meter}},
\end{equation}
and can be described using the following power-law expression:
\begin{equation}
\norm{\alpha_{g,ref} - \alpha_g}_{y=\SI{0.08}{\meter}} = A\epsilon^m,
\end{equation}
where $A$ is a constant and $m$ is the exponent.
For both interface functions, the error follows an approximate first-order decay with the interface width where $m_{\tanh}=0.953$ and $m_{\cos} = 0.896$.

The bubble plume width is computed and tabulated in \cref{tab:plume_width_cos}.
For the simulation with  $\epsilon=0.01$, the bubble plume width is comparable to corresponding the hyperbolic tangent simulation case (\cref{tab:plume_width_tanh}).
The simulation with $\epsilon=0.02$ yields a result with a smaller difference between the two interface functions, but is still below $3\%$.
Overall, it is found that the use of the piece-wise cosine function as the interface kernel improves the bubble plume width in the most diffuse case, decreasing the difference from the reference solution by almost 10\%.
This is due to the lack of asymptotic approach of the phase field $\phi$ to the solid/fluid interface values $\pm 1$ when using the piece-wise cosine, unlike that observed with the hyperbolic tangent kernel function.

\begin{table}
    \centering
    \caption{Bubble plume width at $y = \SI{0.08}{\meter}$ from simulations using piece-wise cosine diffuse-interface.}
    \label{tab:plume_width_cos}
    \begin{tabular}{ccc}
        \hline
        Study & $x_{\text{plume}}$ ($\times \num{e-2}~\si{\meter}$) & Difference (\%) \\
        \hline
        Reference & $3.21$ & -- \\
        $\epsilon = 0.01$ & $3.17$ & $1.25$ \\
        $\epsilon = 0.02$ & $3.13$ & $2.49$ \\
        $\epsilon = 0.04$ & $3.03$ & $5.46$ \\
        $\epsilon = 0.08$ & $2.75$ & $14.2$ \\
        $\epsilon = 0.1$ & $2.54$ & $20.9$ \\
        \hline
    \end{tabular}
\end{table}

\subsection{Flow Past a Cylinder} \label{sec:cylinder}

The diffuse-interface method is used to simulate two-phase flow past a stationary cylinder, a classical benchmark for single-phase flow, but not well studied for dispersed multiphase flows.
Multiphase simulations are performed using both the hyperbolic tangent and piece-wise cosine kernel functions for the diffuse interface functions.
For the hyperbolic tangent case, the cylinder is defined using the following function:
\begin{equation}
\phi(\tilde{\vb*{x}}) = -\tanh(\frac{\norm{\tilde{\vb*{x}} - \tilde{\vb*{x}}_c} - \tilde{R}}{0.5\epsilon}),\label{eq:phi_cylinder_tanh}
\end{equation}
\nomenclature{$\epsilon$}{Diffuse-interface width parameter}%
\nomenclature{$\vb*{x}$}{Position vector}%
\nomenclature{$\tilde{\vb*{x}}_c$}{Scaled diffuse-interface position vector}%
where $\tilde{\vb*{x}}_c = \qty(0,0.8)$ is the scaled diffuse-interface position vector that corresponds to the center of the cylinder, $\tilde{R}=0.1$ is the scaled radius of the cylinder and $\epsilon=0.01$.
The piece-wise cosine interface is defined by:
\begin{equation}
\phi(\tilde{\vb*{x}}) = -\cos(-\pi\min\qty[1, \max\qty(0,\frac{\norm{\tilde{\vb*{x}} - \tilde{\vb*{x}}_c} - \tilde{R} +0.5\eta}{\eta})]),\label{eq:phi_cylinder_cos}
\end{equation}
where $\eta = \epsilon\tanh[-1](0.999)$.
In this geometry, the presence of the use of the diffuse-interface is expected to have a larger impact on the flow profile due to the fact that the immersed cylinder is directly in the path of the gas flow.
The diffuse-interface extends the thickness of the solid boundary, especially for simulations with larger diffuse interface widths, resulting in the effective diameter of the cylinder increasing slightly compared to the reference conformal mesh case.
This is expected to contribute to a deviation of the hydrodynamic behavior of the multiphase flow above some critical diffuse interface width.

\begin{figure}
    \centering
    \begin{subfigure}{0.16\linewidth}
        \includegraphics[width=\linewidth]{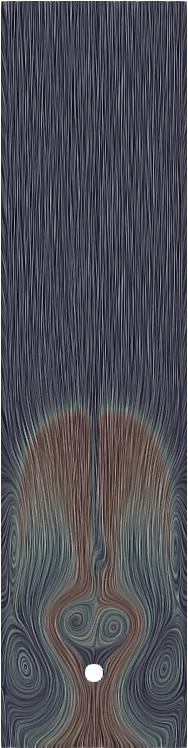}
        \caption*{}
    \end{subfigure}
    \begin{subfigure}{0.16\linewidth}
        \includegraphics[width=\linewidth]{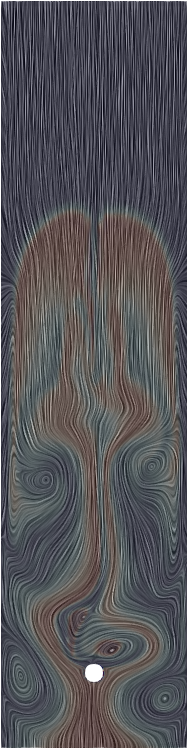}
        \caption*{}
    \end{subfigure}
    \begin{subfigure}{0.16\linewidth}
        \includegraphics[width=\linewidth]{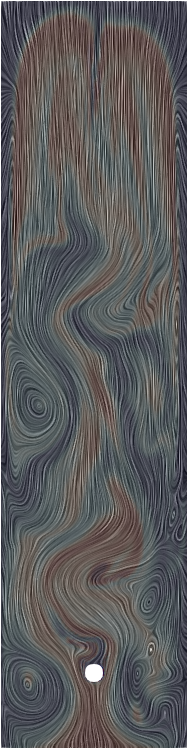}
        \caption*{}
    \end{subfigure}
    \begin{subfigure}{0.112\linewidth}
        \includegraphics[width=\linewidth]{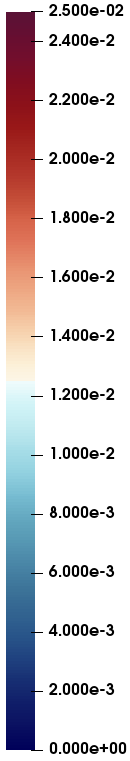}
        \caption*{}
    \end{subfigure}
    \\
    \begin{subfigure}{0.16\linewidth}
        \includegraphics[width=\linewidth]{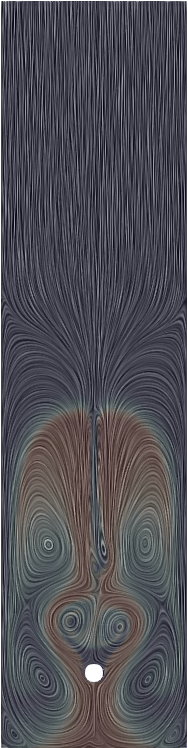}
        \caption*{$\SI{3.13}{\second}$}
    \end{subfigure}
    \begin{subfigure}{0.16\linewidth}
        \includegraphics[width=\linewidth]{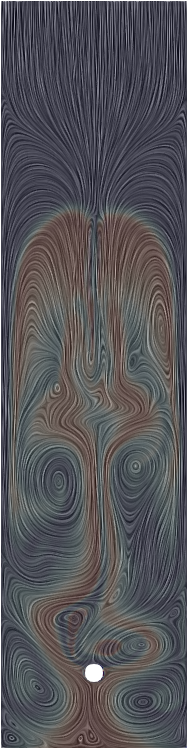}
        \caption*{$\SI{4.69}{\second}$}
    \end{subfigure}
    \begin{subfigure}{0.16\linewidth}
        \includegraphics[width=\linewidth]{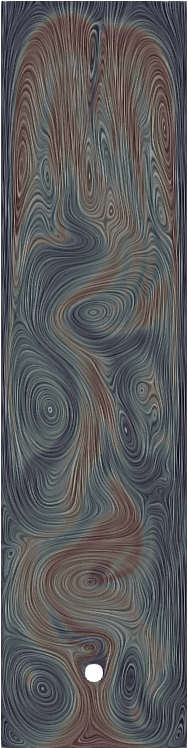}
        \caption*{$\SI{6.25}{\second}$}
    \end{subfigure}
    \begin{subfigure}{0.112\linewidth}
        \includegraphics[width=\linewidth]{TFM_DI/cylinder/cbar}
        \caption*{}
    \end{subfigure}
    \caption{Evolution of gas-liquid flow past a stationary cylinder for the boundary-conformal mesh simulation.
    LICs are of (top) gas and (bottom) liquid phase velocities colored by phase fraction.}
    \label{fig:cylinder_ref}
\end{figure}

\Cref{fig:cylinder_ref} shows the gas and liquid velocity LICs colored with the gas fraction from the reference conformal mesh solution.
In the early stages of the simulation, dispersed gas moves around the cylinder, with a small recirculation region on the upstream side of the cylinder.
Unlike flow through a rectangular channel, gas recirculation is also present in the region near the cylinder.
As the gas travels further up the channel ($t=\SI{3.13}{\second}$), it converges into a single large plume and is convected downstream.
There are two zones of liquid recirculation near the inlet, one on each side of the dispersed gas phase.
The recirculation zone grows in size with time and the vortices begin to detach from their previously stationary location ($t=\SI{3.13}{\second}$).
The flow becomes increasingly unsteady following the initial detachment of vortices, resulting in an undulating column of dispersed gas phase and a distorted bubble plume downstream.
Indicative of unsteady flow, many recirculation zones are simultaneously present on either side of the undulating dispersed gas column.
These convected vortices also increase the dispersion of the gas phase through redirecting undirectional flow.

\begin{figure}
    \centering
    \begin{subfigure}{0.16\linewidth}
        \includegraphics[width=\linewidth]{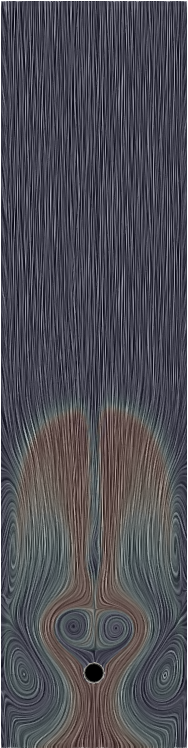}
        \caption*{}
    \end{subfigure}
    \begin{subfigure}{0.16\linewidth}
        \includegraphics[width=\linewidth]{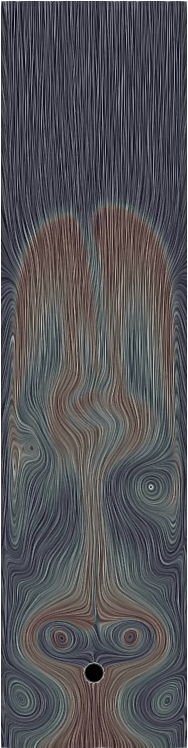}
        \caption*{}
    \end{subfigure}
    \begin{subfigure}{0.16\linewidth}
        \includegraphics[width=\linewidth]{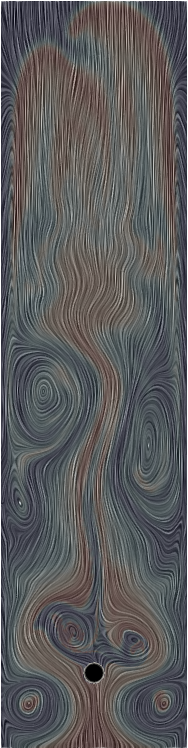}
        \caption*{}
    \end{subfigure}
    \begin{subfigure}{0.112\linewidth}
        \includegraphics[width=\linewidth]{TFM_DI/cylinder/cbar}
        \caption*{}
    \end{subfigure}
    \begin{subfigure}{0.078\linewidth}
        \includegraphics[width=\linewidth]{TFM_DI/phi_cbar}
        \caption*{}
    \end{subfigure}
    \\
    \begin{subfigure}{0.16\linewidth}
        \includegraphics[width=\linewidth]{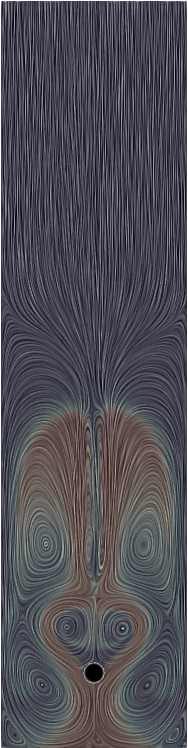}
        \caption*{$\SI{3.13}{\second}$}
    \end{subfigure}
    \begin{subfigure}{0.16\linewidth}
        \includegraphics[width=\linewidth]{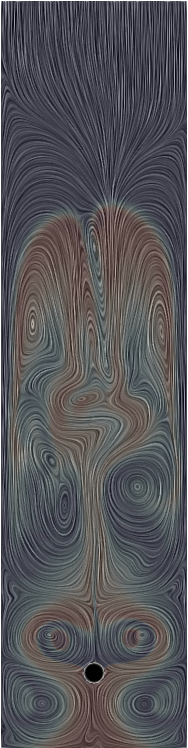}
        \caption*{$\SI{4.69}{\second}$}
    \end{subfigure}
    \begin{subfigure}{0.16\linewidth}
        \includegraphics[width=\linewidth]{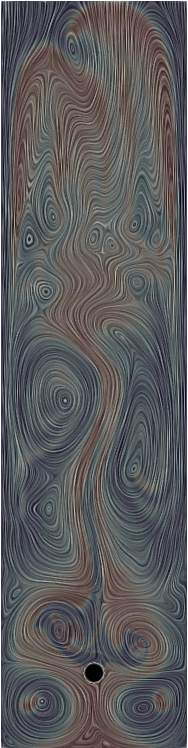}
        \caption*{$\SI{6.25}{\second}$}
    \end{subfigure}
    \begin{subfigure}{0.112\linewidth}
        \includegraphics[width=\linewidth]{TFM_DI/cylinder/cbar}
        \caption*{}
    \end{subfigure}
    \begin{subfigure}{0.078\linewidth}
        \includegraphics[width=\linewidth]{TFM_DI/phi_cbar}
        \caption*{}
    \end{subfigure}
    \caption{Evolution of gas-liquid flow past a stationary cylinder with a hyperbolic tangent diffuse-interface and $\epsilon = 0.01$. The diffuse-interface is in gray-scale and LICs are of (top) gas and (bottom) liquid velocities.}
    \label{fig:cylinder_tanh}
\end{figure}

\begin{figure}
    \centering
    \begin{subfigure}{0.16\linewidth}
        \includegraphics[width=\linewidth]{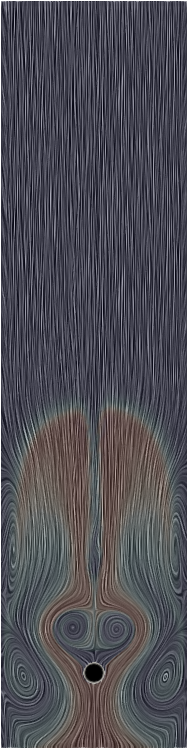}
        \caption*{}
    \end{subfigure}
    \begin{subfigure}{0.16\linewidth}
        \includegraphics[width=\linewidth]{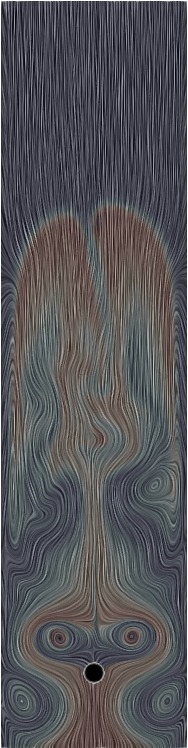}
        \caption*{}
    \end{subfigure}
    \begin{subfigure}{0.16\linewidth}
        \includegraphics[width=\linewidth]{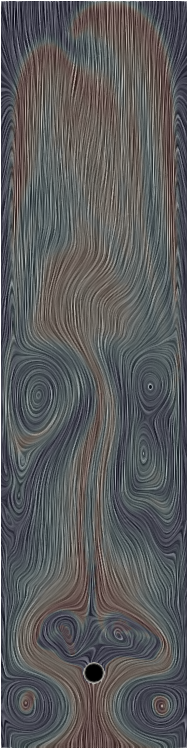}
        \caption*{}
    \end{subfigure}
    \begin{subfigure}{0.112\linewidth}
        \includegraphics[width=\linewidth]{TFM_DI/cylinder/cbar}
        \caption*{}
    \end{subfigure}
    \begin{subfigure}{0.078\linewidth}
        \includegraphics[width=\linewidth]{TFM_DI/phi_cbar}
        \caption*{}
    \end{subfigure}
    \\
    \begin{subfigure}{0.16\linewidth}
        \includegraphics[width=\linewidth]{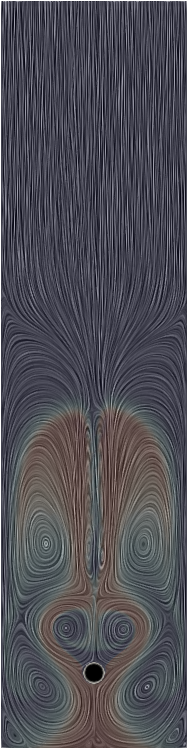}
        \caption*{$\SI{3.13}{\second}$}
    \end{subfigure}
    \begin{subfigure}{0.16\linewidth}
        \includegraphics[width=\linewidth]{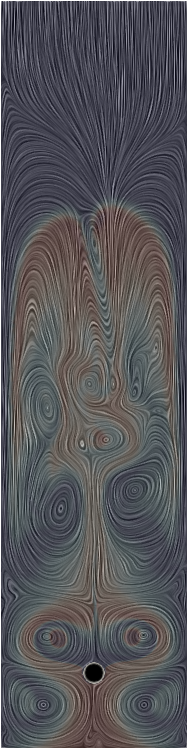}
        \caption*{$\SI{4.69}{\second}$}
    \end{subfigure}
    \begin{subfigure}{0.16\linewidth}
        \includegraphics[width=\linewidth]{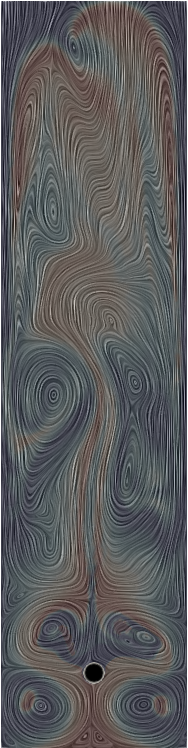}
        \caption*{$\SI{6.25}{\second}$}
    \end{subfigure}
    \begin{subfigure}{0.112\linewidth}
        \includegraphics[width=\linewidth]{TFM_DI/cylinder/cbar}
        \caption*{}
    \end{subfigure}
    \begin{subfigure}{0.078\linewidth}
        \includegraphics[width=\linewidth]{TFM_DI/phi_cbar}
        \caption*{}
    \end{subfigure}
    \caption{Evolution of gas-liquid flow past a stationary cylinder with a piece-wise cosine diffuse-interface and $\epsilon = 0.01$. The diffuse-interface is in gray-scale and LICs are of (top) gas and (bottom) liquid velocities.}
    \label{fig:cylinder_cos}
\end{figure}

\Cref{fig:cylinder_tanh,fig:cylinder_cos} show the simulation results at the same simulation times using the diffuse-interface method with both the (i) hyperbolic tangent and (ii) piece-wise cosine kernels with $\epsilon=0.01$.
The results are not qualitatively different for this diffuse-interface width for either kernel functions.
At $t=\SI{3.13}{\second}$, the gas phase fraction profile and the velocity LICs appear to be the same as the results from the reference simulation for both interface functions.
However, both diffuse-interface simulations deviate starting at $t=\SI{4.69}{\second}$ onward.
The recirculation zones in the wake of the cylinder predicted by the diffuse-interface simulations are wider and closer to the cylinder.
The recirculation zones around the cylinder also appear to be less distorted when compared to the reference case.
This appears to have affected the evolution of the gas and velocity profiles, resulting in similar features but different gas fractions and velocity profiles, confirming the prediction made earlier in this section.

\begin{figure}
    \centering
    \includegraphics[width=0.75\linewidth]{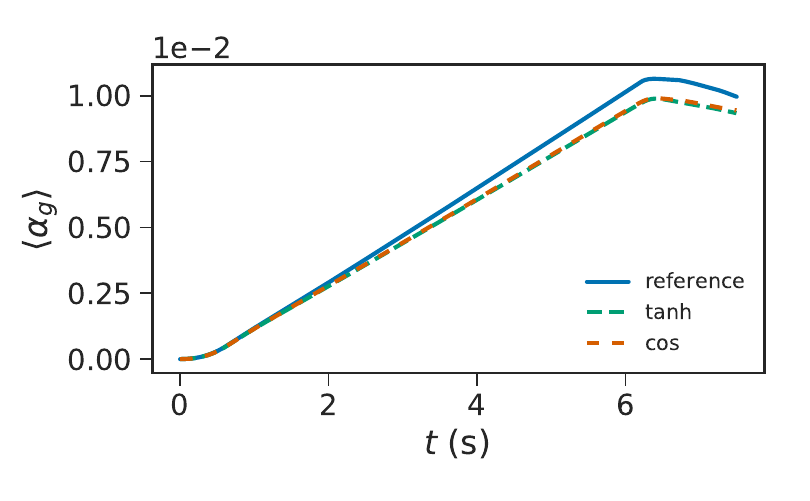}
    \caption{Time evolution of overall gas hold-up in flow past a stationary cylinder.}\label{fig:int_alpha_g_comp_cylinder}
\end{figure}

The time evolution of the overall gas hold-up is shown in \cref{fig:int_alpha_g_comp_cylinder}.
In the early stages of the simulation, the hold-up evolves in the same manner as the reference solution.
The interface function does not appear to significantly affect the solution at $\epsilon=0.01$, supporting the results from the previous subsection.
But as the diffuse-interface interacts with the flow, the gas hold-up diverges from the reference solution.
This corresponds to the observations made in \cref{fig:cylinder_ref,fig:cylinder_tanh,fig:cylinder_cos}.
While the magnitude and the slope of the gas hold-up profiles from the diffuse-interface simulations vary from the reference solution, the qualitative behavior is still the same.

\section{Conclusions} \label{sec:conclusions}

In this work, a novel diffuse solid-fluid interface method is presented for imposing solid boundaries in systems with dispersed multiphase flow conditions.
The dispersed multiphase flow regime is focused on due to its pervasiveness in chemical engineering processes, with the multiphase two-fluid model used for simulations.
No-slip solid/multiphase fluid boundary conditions are imposed through blending the conservation equations of the multiphase two-fluid model with that of a stationary non-deforming solid, resulting in a smooth transition from the solid boundary to the multiphase fluid domain.
To validate the presented method, simulations of channel flow and flow past a cylinder are performed and the results are compared to results from simulations with boundary-conformal meshes.
The results from the diffuse-interface method for simulations of channel flow are found to be in agreement with the reference solution when the diffuse-interface is sufficiently small.
For small diffuse-interface widths, the choice of the interface function has negligible effect on the accuracy of the solution.
Two-phase gas/liquid flow past a stationary cylinder simulations are observed to be in agreement with the reference conformal mesh solution during early stages of the simulation.
However, as the dispersed gas phases is convected towards and around the immersed cylinder, the diffuse-interface representation of the cylinder is found to affect the flow profile and the overall gas hold-up.

The diffuse interface method and simulation observations presented represent a significant step towards the use of immersed boundary-type methods for simulations involving dispersed multiphase flows within complex geometries.
This approach could enable simulation-based design and optimization using multiphase CFD, where evolving the geometry does not require remeshing, along with improved simulation stability resulting from the use of structured meshes.

\section{Background} \label{sec:background}

\subsection{Diffuse Interface Method}

Physical boundaries that are defined using fictitious domain and immersed boundary methods are generally sharp boundaries whose effect may be approximated through the distribution of the boundary over several mesh elements.
This requires the solution field to be interpolated from the physical boundary to the nearest neighboring node/cell \citep{Patel2018}.
The interpolation must be done intermittently throughout the simulation to maintain accuracy and stability.
Special consideration must also be paid when handling mesh elements that are cut by the embedded boundaries \citep{Nguyen2018}.

On the opposite end of the spectrum, the diffuse domain/interface method defines the physical boundaries using a phase-field that approximates the domain boundary by a diffuse region.
Changes in the fluid-solid interface are captured by evolving the phase-field, which does not require interpolation.
For example, the phase-field can vary between zero and one \citep{Nguyen2018}:
\begin{equation}
    \phi = \begin{cases}
        1, \quad \text{physical domain,}\\
        0, \quad \text{otherwise,}
    \end{cases}
\end{equation}
where $\phi$ is the phase-field.
The physical boundary will be represented by the region in which $\phi \in (0,1)$.
The thickness of this region and the transition between the two $\phi$ values are controlled by the function used to define $\phi$.

\subsection{Two-Fluid Model} \label{sec:TFM_eqns}

Dispersed gas-liquid flows are modeled using the two-fluid model, where each phase is considered to be a continuous fluid \citep{Ishii2011}.
The instantaneous behavior of the fluid is averaged over time and phase fractions are used to indicate the spatially-varying composition of the multiphase fluid.
Each of the fluids has its own set of conservation equations and the interactions between the fluids are accounted for through constitutive interphase momentum transfer relationships.
The governing equations of the two-fluid model are given as \citep{Ishii2011}:
\begin{subequations}
    \begin{align}
        \pdv{(\alpha_{q} \rho_{q})}{t} + \div (\alpha_{q}\rho_{q} \vb*{v}_{q})&= 0,\label{eq:continuity}\\
        \begin{split}
            \pdv{\qty(\alpha_{q} \rho_{q} \vb*{v}_{q})}{t} + \div (\alpha_{q} \rho_{q} \vb*{v}_{q}\vb*{v}_{q}) &= -\grad\qty(\alpha_q P_{q}) + \div \qty(\alpha_{q}\vb*{\tau}_{q})+\alpha_{q} \rho_{q} \vb*{g}_{q}\\
            &\qquad  + {\vb*{M}_{q}}  + P_{q,i} \grad \alpha_q - \grad\alpha_q\vdot \vb*{\tau}_{q,i},
        \end{split}\label{eq:momentum_eq}
    \end{align}%
\end{subequations}
\nomenclature{$\alpha$}{Time-averaged local phase fraction}%{--}{--}%
\nomenclature{$\rho$}{Density}%{M L\textsuperscript{-3}}{$\si{\kg~\meter^{-3}}$}%
\nomenclature{$t$}{Time}%{T}{$\si{\second}$}%
\nomenclature{$\vb*{v}$}{Velocity vector}%{L T\textsuperscript{-1}}{$\si{\meter~\second^{-1}}$}%
where $\vb*{v}_q$ is the phasic velocity, $\rho_q$ is the phasic density, $\alpha_q$ is the phase fraction of phase $q$, $P_q$ is the phasic pressure, $\vb*{\tau}_q$ is the phasic viscous stress tensor, $\vb*{g}_q$ is the phasic gravitational force, $\vb*{M}_q$ is the momentum exchange term and the subscript $i$ denote interfacial quantities.

The interphase momentum transfer term may include contributions from various modes of transfer including drag, lift, virtual mass and wall lubrication \citep{Ishii2011,Lahey2001,Antal1991}.
Drag is the largest contributor to the momentum exchange between phases is dispersed flow regimes \citep{Weller2005}.
This is due to the pressure imbalance and shear forces at the gas-liquid interface.
The drag force for the continuous phase, $c$, due to the movement of the dispersed phase, $d$, is given as \citep{Ishii2011}:
\begin{equation}
    {\vb*{M}_{c,drag}} = \frac{1}{2}\rho_c\alpha_d \frac{C_D}{r_d}\norm{\vb*{v}_r}\vb*{v}_r,
    \label{eq:drag_force}
\end{equation}
% Ishii2011 p. 324 eqn 12-31
\nomenclature{$C_D$}{Drag coefficient}%
\nomenclature{$d$}{Diameter}%
\nomenclature{$r_d$}{Volume to projected area ratio}%
where $r_d$ is the ratio of the volume to the projected area of the bubble/particle, $C_D$ is the drag coefficient and $\vb*{v}_r$ is the relative velocity between the dispersed and continuous phases, $\vb*{v}_r = \vb*{v}_d - \vb*{v}_c$.
In spherical bubbles, this becomes:
\begin{equation}
    {\vb*{M}_{c,drag}} =  \frac{3}{4}\rho_c\alpha_d \frac{C_D}{d_d}\norm{\vb*{v}_r}\vb*{v}_r,
    \label{eq:drag_spherical}
\end{equation}
where $d_d$ is the bubble/particle diameter.
The drag force for the dispersed phase is computed using the following property of the interphase momentum exchange:
\begin{equation}
    {\vb*{M}_{c}} = -{\vb*{M}_{d}}.
    \label{eq:mom_trans3}
\end{equation}

In segregated flows, the interfacial shear stress term in \cref{eq:momentum_eq} has a significant effect on the momentum of the fluid \citep{Ishii2011}.
Given that the focus of this work is on the dispersed flow regime, this term is assumed to be negligible.
Additionally, in the dispersed regime, the interfacial pressure of the phases are assumed to be equal \citep{Drew1998,Ishii2011} (\textit{i.e.} $P_{c,i} = P_{d,i} = P_{int}$) and the pressure of the dispersed phase can be approximated by the interfacial pressure ($P_d \approx P_{d,i} = P_{int}$) \citep{Ishii2011}.
The interfacial pressure is approximated by a volume average of the analytical solution of potential flow around a single sphere \citep{Stuhmiller1977,Antal1991}:
\begin{equation}
    P_{c,i} = P_c - C_P \rho_c \vb*{v}_r \vdot \vb*{v}_r,
    \label{eq:interfacial_pressure}
\end{equation}
\nomenclature{$C_{P}$}{Interfacial pressure coefficient}%{--}{--}%
where $C_P$ is the interfacial pressure coefficient.
Thus, the momentum equations in a gas-liquid flow system is given as:
\begin{subequations}
    \begin{align}
        \begin{split}
            \pdv{\qty(\alpha_l \rho_l \vb*{v}_l)}{t} + \div (\alpha_{l} \rho_{l} \vb*{v}_{l}\vb*{v}_{l}) &= -\alpha_l\grad P_{l} + \div \qty(\alpha_{l}\vb*{\tau}_{l})+\alpha_{l} \rho_{l} \vb*{g} \\
            &\quad + \frac{3}{4}\alpha_g\rho_l\frac{C_{D}}{d_b}\norm{\vb*{v}_r}\vb*{v}_r - C_p \vb*{v}_r\vdot\vb*{v}_r \grad \alpha_l,
        \end{split}\label{eq:mom_drag_c} \\
        \begin{split}
            \pdv{\qty(\alpha_g \rho_g \vb*{v}_g)}{t} + \div (\alpha_{g} \rho_{g} \vb*{v}_{g}\vb*{v}_{g}) &= -\alpha_g\grad \qty(P_l - C_p \vb*{v}_r\vdot\vb*{v}_r) + \div \qty(\alpha_{g} \vb*{\tau}_{g})\\
            &\quad +\alpha_{g} \rho_{g} \vb*{g} - \frac{3}{4}\alpha_g\rho_l\frac{C_{D}}{d_b}\norm{\vb*{v}_r}\vb*{v}_r. \end{split}\label{eq:mom_drag_d}
    \end{align}\label{eq:mom_drag}%
\end{subequations}

\section{Methodology}\label{sec:methodology}
The solid physical boundaries are imposed by blending the governing equations of the fluid with the solid Dirichlet boundary conditions.
The diffuse-interface is described by the smooth function $\phi$, whose value is $\pm 1$ inside the phases and is between $(-1,1)$ in the interface region \citep{Shen2010}:
\begin{equation}
    \phi = \begin{cases}
        -1, & \text{fluid}, \\
        1, & \text{solid}.
    \end{cases}\label{eq:phi}%
\end{equation}
From \cref{eq:phi}, the governing equations of the fluid are weighted by $(1-\phi)/2$ to ensure that the equations are active inside the fluid.
Similarly, the solid velocity boundary conditions are weighted by $(1+\phi)/2$ so that the conditions are inactive inside the fluid but active in the solid.
The gradient of the phase-field is the normal vector from the interface and the Neumann boundary condition can be imposed using $\vb*{n} \approx \grad{\phi}/\norm{\grad{\phi}}$.

An example of this diffuse-interface approach is described using the following Poisson problem:
\begin{equation}
    -\laplacian{y} = f \quad \text{on}~\Omega, \qquad \vb*{n}\vdot\grad{y} = h \quad \text{on}~\Gamma_N, \qquad y = g \quad \text{on}~\Gamma_D. \label{eq:poisson_problem}
\end{equation}
The physical domain is denoted by $\phi = -1$ and the area outside the physical domain by $\phi = 1$.
The equation is then weighted by $(1-\phi)/2$ and the Dirichlet condition is weighted by $(1+\phi)/2$:
\begin{equation}
    \frac{1-\phi}{2}\qty(\laplacian{y} + f) + \frac{1+\phi}{2}\qty(y-g) = 0. \label{eq:poisson_blended}
\end{equation}
Taking the inner product of \cref{eq:poisson_blended} with the test function, $\varphi$:
\begin{equation}
    \inp*{\frac{1-\phi}{2}\laplacian{y}}{\varphi}_{\Omega} + \inp*{\frac{1-\phi}{2}f}{\varphi}_{\Omega} + \inp*{\frac{1+\phi}{2}\qty(y-g)}{\varphi}_{\Omega} = 0, \label{eq:poisson_inp}
\end{equation}
where $\inp{\centerdot}{\centerdot}$ is the inner product operator.
The Neumann boundary condition is obtained by applying integration by parts to the Laplacian term:
\begin{equation}
    \inp*{\frac{1-\phi}{2}\laplacian{y}}{\varphi}_{\Omega} = \inp*{\frac{1-\phi}{2}\vb*{n}\vdot\grad{y}}{\varphi}_{\Gamma'_N} + \inp*{\frac{1}{2}\grad{\phi} \vdot \grad{y}}{\varphi}_{\Omega} - \inp*{\frac{1-\phi}{2}\grad{y}}{\grad{\varphi}}_{\Omega},
\end{equation}
where $\Gamma'_N$ is the part of the simulation domain boundary that the Neumann boundary condition applies to and $\vb*{n}$ is the unit normal (outward) of the surface bounding the domain.
Substituting this back into \cref{eq:poisson_inp} and applying the Neumann boundary condition:
\begin{equation}
    \begin{split}
        \inp*{\frac{1-\phi}{2}h}{\varphi}_{\Gamma'_N} + \inp*{\frac{1}{2}h\norm{\grad{\phi}}}{\varphi}_{\Omega} - \inp*{\frac{1-\phi}{2}\grad{y}}{\grad{\varphi}}_{\Omega} \\
        + \inp*{\frac{1-\phi}{2}f}{\varphi}_{\Omega} + \inp*{\frac{1+\phi}{2}\qty(y-g)}{\varphi}_{\Omega} = 0,
    \end{split}\label{eq:poisson_weak_form}%
\end{equation}
where the second term of the left-hand side is obtained from substituting the definition of the diffuse-interface unit normal into the Neumann boundary condition in \cref{eq:poisson_problem}:
\begin{equation}
    \frac{\grad{\phi}}{\norm{\grad{\phi}}}\vdot\grad{y} = h.
\end{equation}
\Cref{eq:poisson_weak_form} is the weak formulation of \cref{eq:poisson_problem} with a diffuse-interface.
The Neumann boundary condition is imposed via the first and/or second terms, depending on the location of the Neumann boundary.
If the Neumann boundary condition is on the simulation domain, the $\frac{1-\phi}{2}h$ term on $\Gamma'_N$ is used to impose the Neumann boundary condition.
However, if the boundary is defined by the diffuse-interface, the $\frac{1}{2}h\norm{\grad{\phi}}$ term in $\Omega$ is used instead.
In the case where the boundary condition applies on both the simulation and diffuse-interface boundaries, then both terms are used.
Similarly, should the Dirichlet boundary condition also apply to parts of the simulation domain boundary, the boundary condition is applied by setting $y=g$ on $\Gamma'_D$.

\subsection{Time Discretization} \label{sec:time_discretization}

Time-integration of the conservation equations is performed using an adaptive second/third order semi-implicit Adams-Bashforth/Backward-Differentiation (AB/BDI23) scheme \citep{Peyret2002}.
The third order AB/BDI3 scheme is used to estimate the local error of the second order scheme.
The explicit terms in the equation are discretized using the Adams-Bashforth scheme and the time derivative is discretized using backward-differentiation \citep{Peyret2002}.
The following notation will be used to denote the numerator of the discretized time derivative:
\begin{align}
    \vb*{v}^{*'} &= a_0 \vb*{v}^* + \sum_{j=1}^{k} a_j \vb*{v}^{n+1-j},\\
    \alpha^{(n+1)'} &= \sum_{j=0}^{k} a_j \alpha^{n+1-j},
\end{align}
\nomenclature{$a_j$}{Backward-differentiation coefficient}%
where $a_j$ is a coefficient associated with backward-differentiation that will later be defined and $k$ is the order of the method.
The discretized explicit terms will be denoted as follows:
\begin{equation}
    f^{n'} = \sum_{j=0}^{k-1}b_j f^{n-j},
\end{equation}
\nomenclature{$b_j$}{Adams-Bashforth coefficient}%
where $b_j$ is a coefficient associated with the Adams-Bashforth scheme.
The procedure to calculate $a_j$ and $b_j$ are outlined in the supplementary material.

\subsection{Diffuse-Interface for Two-Fluid Model Equations} \label{sec:TFM}

In this study, a scaled non-dimensional form of governing equations solved using the phase-bounded incremental pressure correction scheme (IPCS) \citep{Treeratanaphitak2019}.
The scaled equations are scaled using the following dimensionless parameters: $\tilde{\boldsymbol{v}} = \boldsymbol{v}/v_s$, $\tilde{t} = t/t_s$, $\tilde{\boldsymbol{x}} = \boldsymbol{x}/x_s$, $\tilde{P} = (P-P_0)/P_s$,  $\tilde{\boldsymbol{g}} = \boldsymbol{g}/g_s$, $\tilde{\grad} = x_s\grad$ and $ \tilde{d_b} = d_b/x_s$.
This results in the following scaled equations:
\begin{subequations}
    \begin{align}
        \begin{split}  \pdv{\tilde{\vb*{v}}_{l}}{ \tilde{t}} +  \tilde{\vb*{v}}_{l}\vdot\tilde{\grad} \tilde{\vb*{v}}_{l} &= -Eu_l \tilde{\grad} \tilde{P}_l + \frac{1}{Re_l}\frac{\tilde{\grad}\alpha_l \vdot \tilde{\vb*{\tau}}_l}{\alpha_l}  + \frac{1}{Re_l}\tilde{\div}\tilde{\vb*{\tau}}_l+ \frac{1}{Fr^2} \tilde{\vb*{g}} \\
            & \quad +\frac{3}{4}\frac{\alpha_g}{\alpha_l} \frac{C_{D}}{\tilde{d}_b}\norm{\tilde{\vb*{v}}_r}\tilde{\vb*{v}}_r - C_P \tilde{\vb*{v}}_r \vdot\tilde{ \vb*{v}}_r\frac{\tilde{\grad}\alpha_l}{\alpha_l} , \end{split} \label{eq:scaled_mom_l}\\
        \begin{split} \pdv{\tilde{\vb*{v}}_{g}}{\tilde{t}} +   \tilde{\vb*{v}}_{g}\vdot\tilde{\grad} \tilde{\vb*{v}}_{g} &= -Eu_g \tilde{\grad} \qty(\tilde{P}_l - C_P \tilde{\vb*{v}}_r \vdot\tilde{ \vb*{v}}_r\frac{\rho_l}{\rho_g}) + \frac{1}{Re_g}\frac{\tilde{\grad}\alpha_g \vdot \tilde{\vb*{\tau}}_g}{\alpha_g} \\
            & \quad + \frac{1}{Re_g}\tilde{\div}\tilde{\vb*{\tau}}_g+ \frac{1}{Fr^2}  \tilde{\vb*{g}}  -\frac{3}{4}\frac{\rho_l}{\rho_g}\frac{C_{D}}{\tilde{d}_b}\norm{\tilde{\vb*{v}}_r}\tilde{\vb*{v}}_r , \end{split} \label{eq:scaled_mom_g}\\
        \pdv{ \alpha_g}{\tilde{t}} + \tilde{\div}\qty(\alpha_g\tilde{\vb*{v}}_g) &= 0, \label{eq:scaled_mass}\\
        \alpha_l &= 1 - \alpha_g,\label{eq:alpha_l}
    \end{align}\label{eq:TFM_scaled}%
\end{subequations}
\nomenclature{$\Omega$}{Simulation domain}%
\nomenclature{$\Gamma$}{Simulation domain boundary}%
\nomenclature{$\Delta t$}{Time step}%
where the dimensionless groups are defined in \cref{tab:parameters}.
\begin{table}
    \caption{Dimensionless groups}
    \label{tab:parameters}
    \centering
    \begin{tabular}{lc}
        \hline
        Parameter                   &          Expression           \\ \hline
        Time                        &        $t_s = v_s/x_s$        \\
        \multirow{2}{5em}{Pressure} &     $P_s = \rho_l g_s h$      \\
        &           $P_0 = 0$           \\
        Euler number                &   $Eu_q = P_s/\rho_q v_s^2$   \\
        Reynolds number             & $Re_q = \rho_q v_s x_s/\mu_q$ \\
        Froude number               &   $Fr = v_s/\sqrt{g_s x_s}$   \\ \hline
    \end{tabular}
\end{table}
\nomenclature{$Re$}{Reynolds number}%
\nomenclature{$Eu$}{Euler number}%
\nomenclature{$Fr$}{Froude number}%

The diffuse solid-fluid interface is imposed by blending the governing equations of the two-fluid model (\cref{eq:TFM_scaled}) and the solid Dirichlet boundary condition together.
This is achieved by weighting the governing equations and solid boundary condition by $(1-\phi)/2$ and $(1+\phi)/2$, respectively.
The weighting allows for integrals over the physical domain to be reformulated into volume integrals over the simulation domain \citep{Nguyen2018}.
The resulting system of equations is as follows:
\begin{align}
    \begin{split}
        \frac{1-\phi}{2}\qty(\frac{\tilde{\vb*{v}}_{q}^{*'}}{\Delta t} +  \tilde{\vb*{v}}_{q}^{n'}\vdot\tilde{\grad} \tilde{\vb*{v}}_{q}^{n'}) &= \frac{1-\phi}{2}\left[\vb*{RHS}_q^n + \frac{1}{Re_q}\frac{\tilde{\grad}\alpha_q^{n'} \vdot \tilde{\vb*{\tau}}_q^{n+1}}{\alpha_q^{n'}}\right.\\
        & \qquad \qquad \quad  \left.+ \frac{1}{Re_q}\tilde{\div}\tilde{\vb*{\tau}}_q^{n+1}+ \frac{1}{Fr^2} \tilde{\vb*{g}} \right]
    \end{split}& & \text{in} \quad\Omega,\label{eq:mom_phi}
\end{align}
where:
\begin{subequations}
    \begin{align}
        \vb*{RHS}_l^n &= -Eu_l \tilde{\grad} \tilde{P}^{n}_l+ \frac{3}{4}\frac{\alpha_g^{n'}}{\alpha_l^{n'}} \frac{C_{D}}{\tilde{d}_b}\norm{\tilde{\vb*{v}}_r^{n'}}\tilde{\vb*{v}}_r^{n'}- C_P \tilde{\vb*{v}}_r^{n'} \vdot\tilde{ \vb*{v}}_r^{n'}\frac{\tilde{\grad}\alpha_l^{n'}}{\alpha_l^{n'}}, \label{eq:RHS_l}\\
        \vb*{RHS}_g^n &= -Eu_g \tilde{\grad} \qty(\tilde{P}_l^{n} - C_P \tilde{\vb*{v}}_r^{n'} \vdot\tilde{ \vb*{v}}_r^{n'}\frac{\rho_l}{\rho_g}) -\frac{3}{4}\frac{\rho_l}{\rho_g}\frac{C_{D}}{\tilde{d}_b}\norm{\tilde{\vb*{v}}_r^{n'}}\tilde{\vb*{v}}_r^{n'},  \label{eq:RHS_g}
    \end{align}%
\end{subequations}
with the following boundary conditions:
\begin{subequations}
    \begin{align}
        \frac{1+\phi}{2}\tilde{\vb*{v}}_q^* &= \vb*{0}& & \text{in} \quad\Omega, \\
        \frac{1-\phi}{2}\tilde{\vb*{v}}_q^* &=\frac{1-\phi}{2} \tilde{\vb*{v}}_{q,BC}^{n+1} & & \text{on} \quad\Gamma'_D, \\
        \vb*{n}\vdot\frac{1-\phi}{2}\tilde{\vb*{\tau}}_q^{n+1} &= \vb*{0} & & \text{on} \quad\Gamma'_N.
    \end{align}%
\end{subequations}
The time discretization follows the notation defined in the previous section.
The weak formulation of \cref{eq:mom_phi} follows the same procedure as the Poisson equation example outlined earlier in the section but with the two-fluid model equations.
Taking the inner product of \cref{eq:mom_phi} and the test function yields:
\begin{equation}
    \begin{split}
        \inp*{\frac{1-\phi}{2} \frac{\tilde{\vb*{v}}_{q}^{*'}}{\Delta t}}{\vb*{\varphi}_q}_{\Omega} &+ \inp*{\frac{1+\phi}{2} \frac{a_0\tilde{\vb*{v}}_{q}^{*}}{\Delta t}}{\vb*{\varphi}_q}_{\Omega} \\
        & \quad = -\inp*{\frac{1-\phi}{2} \tilde{\vb*{v}}_{q}^{n'}\vdot\tilde{\grad} \tilde{\vb*{v}}_{q}^{n'}}{\vb*{\varphi}_q}_{\Omega} + \inp*{\frac{1-\phi}{2} \vb*{RHS}_q^n}{\vb*{\varphi}_q}_{\Omega} \\
        & \quad + \inp*{\frac{1-\phi}{2}  \frac{1}{Re_q}\frac{\tilde{\grad}\alpha_q^{n'} \vdot \tilde{\vb*{\tau}}_q^{n+1}}{\alpha_q^{n'}}}{\vb*{\varphi}_q}_{\Omega}  + \inp*{\frac{1-\phi}{2}\frac{1}{Re_q}\tilde{\div}\tilde{ \vb*{\tau}}_q^{*}}{\vb*{\varphi}_q}_{\Omega} \\
        & \quad  + \inp*{\frac{1-\phi}{2} \frac{1}{Fr^2} \tilde{\vb*{g}}}{\vb*{\varphi}_q}_{\Omega}.
    \end{split}\label{eq:TFM_momentum_phi_a}
\end{equation}
The Neumann boundary condition for \cref{eq:mom_phi} is obtained from using integration by parts on the $\frac{1-\phi}{2}\frac{1}{Re_q}\tilde{\div}\tilde{\vb*{\tau}}_q^{*}$ term:
\begin{equation}
    \begin{split}
        \inp*{\frac{1-\phi}{2}\frac{1}{Re_q}\tilde{\div}\tilde{ \vb*{\tau}}_q^{*}}{\vb*{\varphi}_q}_{\Omega} = &\inp*{\frac{1-\phi}{2}\frac{1}{Re_q} \vb*{n} \vdot \tilde{ \vb*{\tau}}_q^{n+1}}{\vb*{\varphi}_q}_{\Gamma'_N} + \inp*{\frac{1}{2 Re_q} \tilde{ \grad} \phi \vdot \tilde{ \vb*{\tau}}_q^{*}}{\vb*{\varphi}_q}_{\Omega} \\
        & - \inp*{\frac{1-\phi}{2}\frac{1}{Re_q} \tilde{ \vb*{\tau}}_q^{*}}{ \tilde{ \grad}\vb*{\varphi}_q}_{\Omega}.
    \end{split}\label{eq:tau_phi}
\end{equation}
\nomenclature{$\inp{\centerdot}{\centerdot}$}{Inner product}%
The second term in the right-hand side of \cref{eq:tau_phi} allows for the imposition of a Neumann boundary condition at the solid-fluid interface.
In this work, the boundary condition at the solid-fluid interface is a Dirichlet boundary condition and the term is therefore left unconstrained.
The weak formulation is thus:
\begin{equation}
    \begin{split}
        \inp*{\frac{1-\phi}{2} \frac{\tilde{\vb*{v}}_{q}^{*'}}{\Delta t}}{\vb*{\varphi}_q}_{\Omega} &+ \inp*{\frac{1+\phi}{2} \frac{a_0\tilde{\vb*{v}}_{q}^{*}}{\Delta t}}{\vb*{\varphi}_q}_{\Omega} \\
        & \quad = -\inp*{\frac{1-\phi}{2} \tilde{\vb*{v}}_{q}^{n'}\vdot\tilde{\grad} \tilde{\vb*{v}}_{q}^{n'}}{\vb*{\varphi}_q}_{\Omega} + \inp*{\frac{1-\phi}{2} \vb*{RHS}_q^n}{\vb*{\varphi}_q}_{\Omega} \\
        & \quad + \inp*{\frac{1-\phi}{2}  \frac{1}{Re_q}\frac{\tilde{\grad}\alpha_q^{n'} \vdot \tilde{\vb*{\tau}}_q^{n+1}}{\alpha_q^{n'}}}{\vb*{\varphi}_q}_{\Omega}  + \inp*{\frac{1-\phi}{2}\frac{1}{Re_q} \vb*{n} \vdot \tilde{ \vb*{\tau}}_q^{n+1}}{\vb*{\varphi}_q}_{\Gamma'_N} \\
        & \quad - \inp*{\frac{1-\phi}{2}\frac{1}{Re_q} \tilde{ \vb*{\tau}}_q^{*}}{ \tilde{ \grad}\vb*{\varphi}_q}_{\Omega} + \inp*{\frac{1-\phi}{2} \frac{1}{Fr^2} \tilde{\vb*{g}}}{\vb*{\varphi}_q}_{\Omega},
    \end{split}\label{eq:TFM_momentum_phi}
\end{equation}
where the solid boundary condition is weighted by $a_0/\Delta t$ for consistency.
The pressure Poisson equation is derived from \cref{eq:mom_phi} by taking the difference between the weighted momentum equation for $\tilde{ \vb*{v}}_q^{n+1}$ and $\tilde{ \vb*{v}}_q^*$ and neglecting the contributions of convection, viscous stress and interphase momentum transfer:
\begin{equation}
    -\tilde{\div}\qty[\frac{1-\phi}{2}\sum_q Eu_q \alpha_q^{n'} \tilde{ \grad}\qty(\tilde{P}_l^{n+1} - \tilde{P}_l^n)] = \frac{a_0}{\Delta t} \tilde{\div}\qty[\frac{1-\phi}{2}\sum_q\qty( \alpha_q^{n+1}\tilde{ \vb*{v}}_q^{n+1} - \alpha_q^n\tilde{ \vb*{v}}_q^*)].
\end{equation}
The right-hand side term can be separated into two terms:
\begin{equation}
    \begin{split}
        \frac{a_0}{\Delta t} \tilde{\div}\qty[\frac{1-\phi}{2}\sum_q\qty( \alpha_q^{n+1}\tilde{ \vb*{v}}_q^{n+1} - \alpha_q^n\tilde{ \vb*{v}}_q^*)] = &-\frac{a_0}{2\Delta t}\tilde{ \grad}\phi\vdot \sum_q \qty( \alpha_q^{n+1}\tilde{ \vb*{v}}_q^{n+1} - \alpha_q^n\tilde{ \vb*{v}}_q^*) \\
        & +  \frac{ a_0}{\Delta t} \frac{1-\phi}{2}\tilde{\div} \sum_q\qty(\alpha_q^{n+1}\tilde{ \vb*{v}}_q^{n+1} - \alpha_q^n\tilde{ \vb*{v}}_q^*).
    \end{split}
\end{equation}
The first term is only active at the solid-fluid interface and given that the phase fraction and velocity of the solid are always known, this term is assumed to be negligible.
Using the incompressibility condition for the two-fluid model, $\div{\sum_q \alpha_q \vb*{v}_q} = 0$, the pressure Poisson equation for two-phase flow using the diffuse-interface method is thus:
\begin{equation}
    \tilde{\div}\qty[\frac{1-\phi}{2}\sum_q Eu_q \alpha_q^{n'}\tilde{ \grad}\qty(\tilde{P}_l^{n+1} - \tilde{P}_l^n)] = \frac{ a_0}{\Delta t}\frac{1-\phi}{2} \tilde{\div}\qty(\sum_q\ \alpha_q^n\tilde{ \vb*{v}}_q^*),
\end{equation}
with the following weak formulation obtained using integration by parts:
\begin{equation}
    \begin{split}
        \inp*{\frac{1-\phi}{2}\sum_q Eu_q \alpha_q^{n'} \vb*{n}\vdot \tilde{ \grad}\qty(\tilde{P}_l^{n+1} - \tilde{P}_l^n)}{\varphi_p}_{\Gamma'_D} &- \inp*{\frac{1-\phi}{2}\sum_q Eu_q \alpha_q^{n'}  \tilde{ \grad}\qty(\tilde{P}_l^{n+1} - \tilde{P}_l^n)}{\tilde{\varphi}_p}_{\Omega} \\
        & \quad = \inp*{\frac{a_0}{\Delta t} \frac{1-\phi}{2} \tilde{\div}\qty(\sum_q\alpha_q^{n'}\tilde{\vb*{v}}_q^{*})}{\varphi_p}_{\Omega}.
    \end{split}\label{eq:TFM_PP_phi}
\end{equation}
The new velocity update equation is simply sum of the update equation from IPCS weighted by $(1-\phi)/2$ and the solid Dirichlet boundary condition weighted by $(1+\phi)/2$:
\begin{equation}
    \inp*{\frac{1-\phi}{2}a_0 \frac{\tilde{ \vb*{v}}^{n+1}_q - \tilde{ \vb*{v}}^*_q}{\Delta t}}{\vb*{\varphi}_q}_{\Omega}  + \inp*{\frac{1+\phi}{2}\frac{a_0\tilde{ \vb*{v}}^{n+1}_q}{\Delta t}}{\vb*{\varphi}_q}_{\Omega} = -\inp*{\frac{1-\phi}{2}Eu_q\tilde{ \grad}\qty(\tilde{P}_l^{n+1} - \tilde{P}_l^n)}{\vb*{\varphi}_q}_{\Omega}.\label{eq:TFM_update_phi}
\end{equation}
The boundary condition for the gas fraction, $\alpha_g$, at the solid-fluid interface is $\alpha_g = 0$ (liquid wets the wall).
Using the same blending procedure to apply the boundary condition yields the following:
\begin{equation}
    \inp*{\frac{1-\phi}{2}\frac{\alpha_g^{(n+1)'}}{\Delta t}}{\varphi_\alpha}_{\Omega} + \inp*{\frac{1+\phi}{2}\frac{a_0\alpha_g^{n+1}}{\Delta t}}{\varphi_\alpha}_{\Omega} + \inp*{\frac{1-\phi}{2}\tilde{\div}\qty(\alpha_g^{n+1}\tilde{\vb*{v}}_g^{n+1})}{\varphi_\alpha}_{\Omega} = 0.
    \label{eq:TFM_alpha_phi}
\end{equation}

\subsection{Simulation Conditions}\label{sec:simulation_conditions}

\begin{figure}
    \centering
    \includegraphics[width=0.4\textwidth]{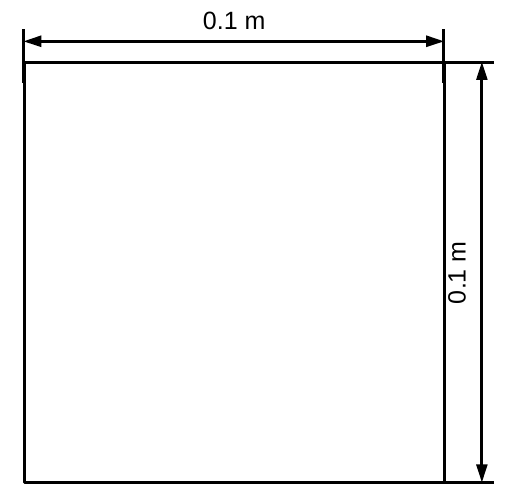}
    \caption{Simulation domain for gas-liquid flow inside a channel with the diffuse-interface method.}
    \label{fig:TFMDI_simulation_domain_channel}
\end{figure}

\begin{figure}
    \centering
    \includegraphics[width=0.25\textwidth]{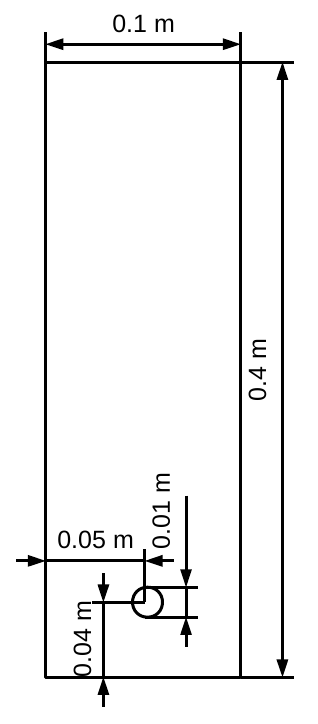}
    \caption{Simulation domain for gas-liquid flow past a stationary cylinder with the diffuse-interface method.}
    \label{fig:TFMDI_simulation_domain_cylinder}
\end{figure}

The diffuse-interface method is used to impose boundary conditions in dispersed gas-liquid simulations of a two-dimensional channel (\cref{fig:TFMDI_simulation_domain_channel}) and flow past a stationary cylinder (\cref{fig:TFMDI_simulation_domain_cylinder}).
The physical properties of the fluids are reported in \cref{tab:phys_prop}.
For the two-dimensional channel case, results from previous work by the authors \citep{Treeratanaphitak2019} are used as the reference case with a conformal mesh.
The width of the channel in \cref{fig:TFMDI_simulation_domain_channel} is twice that of the simulation domain in \citet{Treeratanaphitak2019}.
The channel walls will be imposed using a phase-field and the remaining boundary conditions are the same as in \citet{Treeratanaphitak2019}.
The new inlet boundary conditions are given in \cref{tab:bcs_GL_channel}.
For the case of flow past a cylinder, parabolic velocity and gas fraction profiles are used at the inlet (\cref{tab:bcs_GL_cylinder}), no-slip and zero gas fraction conditions are imposed at the channel and cylinder walls and outflow conditions are used at the outlet.
The simulations are performed with 16 cores (Intel E5-2683 v4 Broadwell 2.1GHz) for approximately two weeks of wall-time using compute notes provided by the Digital Research Alliance of Canada.

\begin{table}
    \centering
    \caption{Physical properties}
    \label{tab:phys_prop}
    \begin{tabular}{lc}
        \hline
        Property & Value \\
        \hline
        Gas density ($\si{\kg\per\meter^3}$) & $10$ \\
        Liquid density ($\si{\kg\per\meter^3}$) & $1000$ \\
        Gas viscosity ($\si{\pascal \second}$) & $2\times 10^{-5}$ \\
        Liquid viscosity ($\si{\pascal \second}$) & $5\times 10^{-3}$ \\
        Bubble diameter ($\si{\meter}$) & $10^{-3}$ \\
        Drag constant & $\max\qty[\frac{24}{Re}\qty(1+0.15Re^{0.687}),0.44], Re = \frac{\rho_l \norm{\vb*{v}_r}d_b}{\mu_l}$ \\
        \hline
    \end{tabular}
\end{table}

\begin{table}
    \centering
    \caption{Initial and inlet conditions for gas-liquid channel flow with diffuse-interface.}
    \label{tab:bcs_GL_channel}
    \begin{tabular}{lc}
        \hline
        & Condition \\
        \hline
        \multirow{3}{*}{Initial} & $\alpha_g\qty(\vb*{x},0) = 0$ \\
        & $\vb*{v}_g\qty(\vb*{x},0) = \vb*{v}_l\qty(\vb*{x},0) = \vb{0}$ \\
        & $P\qty(\vb*{x}, 0) = \rho_l g_s (0.1 - y)$ \\
        \hline
        \multirow{4}{*}{Inlet} & $\vb*{v}_g\qty(x,0,t) = \qty(0, \min\qty(\frac{t}{t_0}, 1)\frac{1-\phi}{2}0.0616\exp\qty[-\frac{\qty(\frac{x}{0.025})^2}{2\sigma^2}]), t_0 = 0.625~\si{\second}, \sigma = 0.1$ \\
        & $\vb*{v}_l\qty(x,0,0) = \vb*{0} $ \\
        &  $\alpha_{g}\qty(x,0,t) = \min\qty(\frac{t}{t_0}, 1)\frac{1-\phi}{2}0.026\exp\qty[-\frac{\qty(\frac{x}{0.025})^2}{2\sigma^2}], t_0 = 0.625~\si{\second}, \sigma = 0.1$ \\
        & $\vb*{n}\vdot\frac{1-\phi}{2}\grad{\qty(P_l\qty(x,0,t) - P_l\qty(x,0,t-\Delta t))} = 0$ \\
        \hline
    \end{tabular}
\end{table}

\begin{table}
    \centering
    \caption{Initial and inlet conditions for gas-liquid flow past a cylinder.}
    \label{tab:bcs_GL_cylinder}
    \begin{tabular}{lc}
        \hline
        & Condition \\
        \hline
        \multirow{3}{*}{Initial} & $\alpha_g\qty(\vb*{x},0) = 0$ \\
        & $\vb*{v}_g\qty(\vb*{x},0) = \vb*{v}_l\qty(\vb*{x},0) = \vb{0}$ \\
        & $P\qty(\vb*{x}, 0) = \rho_l g_s (0.4 - y)$ \\
        \hline
        \multirow{4}{*}{Inlet} & $\vb*{v}_g\qty(x,0,t) = \qty(0, \min\qty(\frac{t}{t_0}, 1) 0.0616\qty(0.025-x^2)), t_0 = 0.625~\si{\second}$ \\
        & $\vb*{v}_l\qty(x,0,0) = \vb*{0} $ \\
        &  $\alpha_{g}\qty(x,0,t) = \min\qty(\frac{t}{t_0}, 1) 0.02\qty(0.025-x^2), t_0 = 0.625~\si{\second}$ \\
        & $\vb*{n}\vdot\grad{P}\qty(x,0,t) = 0$ \\
        \hline
    \end{tabular}
\end{table}

\begin{acknowledgement}
    This research was supported by the Natural Sciences and Engineering Research Council (NSERC) of Canada and the Digital Research Alliance of Canada.
\end{acknowledgement}

\begin{suppinfo}
\begin{itemize}
    \item Description of variable step size coefficients used in the time integration schemes
\end{itemize}

\end{suppinfo}

% uncomment this when submitting
%\processdelayedfloats

\section*{Author Information}
\paragraph{Tanyakarn Treeratanaphitak (corresponding author):} School of Integrated Science and Innovation, Sirindhorn International Institute of Technology, Thammasat University, 99 Moo 18 Paholyothin Road, Klong Nueng, Klong Luang, Pathum Thani 12121, Thailand

Email: \texttt{tanyakarn@siit.tu.ac.th}

\paragraph{Nasser Mohieddin Abukhdeir:} Department of Chemical Engineering, University of Waterloo, 200 University Avenue West, Waterloo, N2L 3G1, ON, Canada

Department of Physics \& Astronomy, University of Waterloo, 200 University Avenue West, Waterloo, N2L 3G1, ON, Canada

Email: \texttt{nmabukhdeir@uwaterloo.ca}

\bibliography{multiphase,computational}

\newpage

\Large {For Table of Contents use only}

\vspace{2em}

\centering
\includegraphics[scale=1]{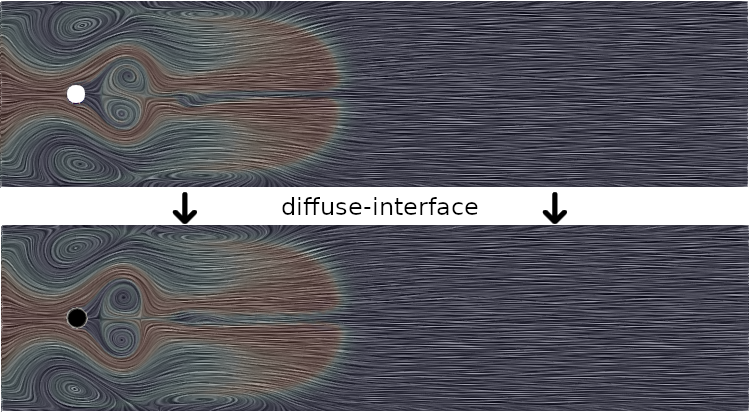}

\end{document}